\documentclass[3p]{elsarticle}
\usepackage[numbers]{natbib}

\usepackage{amssymb,amsmath} 
\usepackage[english,francais]{babel}

\usepackage{graphicx,color,float}
\usepackage{subfigure}

 \usepackage[usenames,dvipsnames]{pstricks}
 \usepackage{epsfig}
 \usepackage{pst-grad} 
 \usepackage{pst-plot} 

\renewcommand{\theequation}{\arabic{equation}}
\def\og{\leavevmode\raise.3ex\hbox{$\scriptscriptstyle\langle\!\langle$~}}
\def\fg{\leavevmode\raise.3ex\hbox{~$\!\scriptscriptstyle\,\rangle\!\rangle$}}

\def\eeq{\relax}
\def\beq#1#2\eeq{\begin{equation}\label{#1}#2\end{equation}}
\def\bal#1#2\eal{\begin{align}\label{#1}#2\end{align}}
\def\bse#1#2\ese{\begin{subequations}\label{#1}#2\end{subequations}}
\def\ba{\begin{aligned}}
\def\ea{\end{aligned}}

\def\det{\operatorname{det}}
\def\dd{\operatorname{d}}
\def\tr{\operatorname{tr}}

\def\div{\operatorname{div}}

\def\a{b}

\renewcommand{\appendix}{
  \setcounter{section}{0}\renewcommand{\thesection}{\Alph{section}}
  \section*{Appendix} }

\def\Appendix#1{
  \setcounter{equation}{0}
  \renewcommand{\theequation}{\thesection.\arabic{equation}}
  \section{#1}  }

\journal{Int. J. Engng. Sci.}

\begin{document}

\begin{frontmatter}
\selectlanguage{english}
\title{Enhanced  acoustic transmission through a slanted grating}  

\author[authorlabel1]{A. N. Norris},
\ead{norris@rutgers.edu}
\author[authorlabel1]{Xiaoshi Su}
\ead{xiaoshi.su@rutgers.edu}


\address[authorlabel1]{ Mechanical and Aerospace Engineering, \\
	Rutgers University, Piscataway NJ 08854-8058, USA }

\begin{abstract}

It is known that an acoustic wave incident on an infinite array of aligned rectangular blocks of a different acoustic material  exhibits total transmission  if certain conditions are met \cite{Maurel13} which relate the unique "intromission" angle of incidence with  geometric and material properties of the slab. This extraordinary acoustic transmission phenomenon holds for any slab thickness, making it analogous to a Brewster effect in optics, and is independent of  frequency as long as the slab microstructure is sub-wavelength in the length-wise direction.  Here we show  that the enhanced transmission  effect is obtained in a slab with  grating elements oriented  obliquely to the slab normal.  The dependence of the intromission angle $\theta_i$ is given explicitly in terms of the orientation angle.   Total transmission is achieved at incidence angles $\pm \theta_i$, with  a relative phase shift between the transmitted amplitudes of the $+\theta_i$ and $- \theta_i$ cases.   These effects are shown to follow from explicit formulas for the transmission coefficient.  In the case of  grating elements that  are 
rigid the results have direct physical interpretation.   The analytical findings are illustrated with  full wave simulations. 
\end{abstract}
\end{frontmatter}

\section{Introduction}\label{sec1}

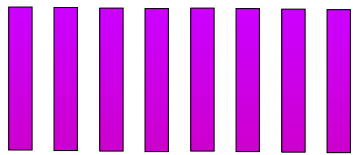
\begin{figure}[h] 
	\centering
\scalebox{.35} 
{
\begin{pspicture}(0,-3.7965624)(14.055,3.7965624)
\definecolor{color245g}{rgb}{0.8,0.0,1.0}
\definecolor{color245f}{rgb}{0.8,0.0,0.8}
\psframe[linewidth=0.002,dimen=outer,fillstyle=gradient,gradlines=2000,gradbegin=color245g,gradend=color245f,gradmidpoint=1.0](2.095,3.113125)(1.215,-2.306875)
\psframe[linewidth=0.002,dimen=outer,fillstyle=gradient,gradlines=2000,gradbegin=color245g,gradend=color245f,gradmidpoint=1.0](3.795,3.093125)(2.915,-2.326875)
\psframe[linewidth=0.002,dimen=outer,fillstyle=gradient,gradlines=2000,gradbegin=color245g,gradend=color245f,gradmidpoint=1.0](5.515,3.073125)(4.635,-2.346875)
\psframe[linewidth=0.002,dimen=outer,fillstyle=gradient,gradlines=2000,gradbegin=color245g,gradend=color245f,gradmidpoint=1.0](7.215,3.053125)(6.335,-2.366875)
\psframe[linewidth=0.002,dimen=outer,fillstyle=gradient,gradlines=2000,gradbegin=color245g,gradend=color245f,gradmidpoint=1.0](8.935,3.073125)(8.055,-2.346875)
\psframe[linewidth=0.002,dimen=outer,fillstyle=gradient,gradlines=2000,gradbegin=color245g,gradend=color245f,gradmidpoint=1.0](10.635,3.053125)(9.755,-2.366875)
\psframe[linewidth=0.002,dimen=outer,fillstyle=gradient,gradlines=2000,gradbegin=color245g,gradend=color245f,gradmidpoint=1.0](12.355,3.033125)(11.475,-2.386875)
\psframe[linewidth=0.002,dimen=outer,fillstyle=gradient,gradlines=2000,gradbegin=color245g,gradend=color245f,gradmidpoint=1.0](14.055,3.013125)(13.175,-2.406875)
\end{pspicture} 
}
	\caption{A section of an infinite  single-layer grating of rigid blocks in a fluid.  
	}
	\label{fig1}
\end{figure}

\begin{figure}[H] 
	\centering
\includegraphics[width=0.75\textwidth]{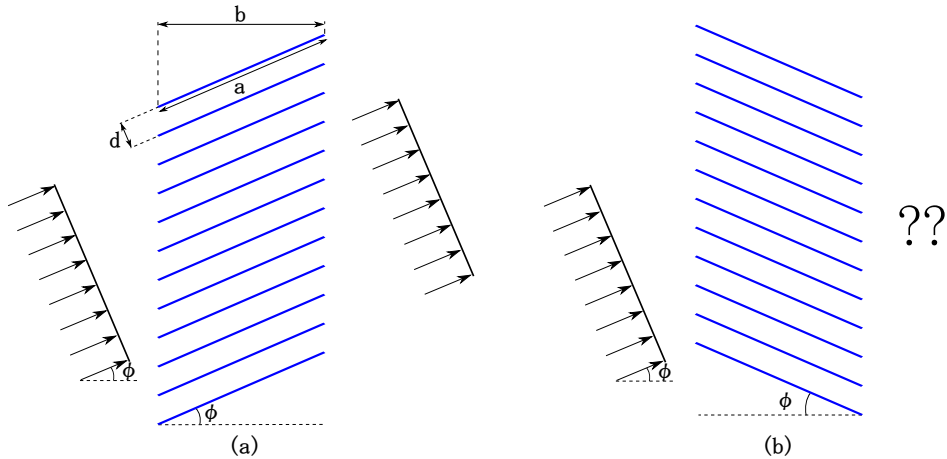}
	\caption{Figure (a) shows a grating of thin rigid strips aligned with the incident plane wave, resulting in perfect transmission.  What will happen if we reverse the orientation  of the slab elements as in (b)?}
	\label{line}
\end{figure}

Consider a slab comprised of rigid rectangles arranged periodically to form a comb-like grating of infinite extent as depicted in Figure \ref{fig1}.  
\citet{D'Aguanno12} showed that such a "single layer grating" (SLG) with rigid filling fraction $f$   exhibits total transmission for  an acoustic plane wave incident at  
\emph{intromission angle} $\theta_i$ satisfying $\cos\theta_i =1-f$.  The  angle is defined relative to the slab normal.  For instance,  the intromission angle is zero 
  for a grating of vanishingly thin rigid plates, $f=0+$;  this  limiting case of $\theta_i = 0$ is intuitively obvious because the acoustic wave does not interact with an infinitesimally thin rigid plate aligned with the acoustic particle 
motion (even the diffraction effects vanish because the diffraction coefficient for parallel incidence on a semi-infinite rigid strip is zero \cite{KELLER1962}). 
Now consider the same thin rigid plates rotated through angle $\phi$ as shown in Figure \ref{line}(a). 
This oblique grating again "obviously" has intromission angle $\theta_i =\phi$, just like the orthogonal grating.     

Now consider the same grating but with the orientation of the plates in the grating reversed while the incident wave remains the same,  Figure \ref{line}(b).    
Remarkably, the two gratings in  Figure \ref{line} are identical in terms of the magnitudes of the reflection and transmission coefficients for all angles of incidence and for all  frequencies for which the homogenization approximations apply.  This equivalence becomes apparent when one realizes that the two gratings present the same effective acoustic impedance in the homogenization limit.  There is, however,  a phase difference between the transmitted waves for the two cases in Figures \ref{line} (a) and (b).  These effects, including the cases in Figure \ref{line}, are derived in this paper in the context of a general SLG composed of rotated elements.   


Extraordinary optical transmission (EOT) through metallic gratings can occur when the openings resonate in Fabry-Perot mode,  which is well established although  very narrow band effect.   Broadband EOT, spanning from DC upwards, has been recently  proposed \cite{Alu2011} and realized \cite{Argyropoulos2012} based on a Brewster angle effect that results from the equivalent long-wavelength properties of the grating. 
\citet{Abek2012} demonstrated Brewster-like broadband extraordinary optical transmission in a thick metal plate with  slits as narrow  as $\lambda$/750.  They showed that   an order of magnitude larger transmission is obtained for  very narrow slits compared to the normal-incidence Fabry-P\'erot resonance transmission peaks.  EOT has also been confirmed experimentally  for TE and TM waves through subwavelength dielectric gratings in the microwave regime \cite{Akarid2014}.

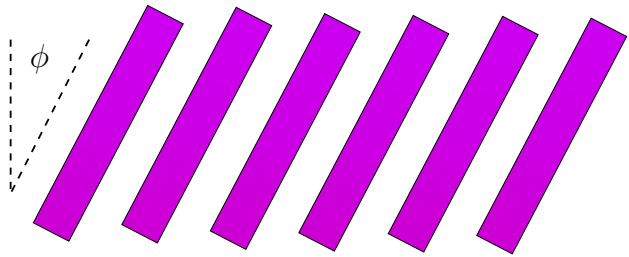
\begin{figure}[h] 
	\centering
\scalebox{.6} 
{
\begin{pspicture}(0,-3.5328372)(14.1,3.492488)
\definecolor{color249g}{rgb}{0.8,0.0,1.0}
\definecolor{color249f}{rgb}{0.8,0.0,0.8}
\rput{-27.565233}(-0.16747421,1.0909739){\psframe[linewidth=0.002,dimen=outer,fillstyle=gradient,gradlines=2000,gradbegin=color249g,gradend=color249f,gradmidpoint=1.0](2.58,3.5968504)(1.7,-1.8231496)}
\rput{-27.565233}(0.07125611,1.9841841){\psframe[linewidth=0.002,dimen=outer,fillstyle=gradient,gradlines=2000,gradbegin=color249g,gradend=color249f,gradmidpoint=1.0](4.52,3.5568504)(3.64,-1.8631496)}
\rput{-27.565233}(0.35626227,2.8660429){\psframe[linewidth=0.002,dimen=outer,fillstyle=gradient,gradlines=2000,gradbegin=color249g,gradend=color249f,gradmidpoint=1.0](6.46,3.4168503)(5.58,-2.0031495)}
\rput{-27.565233}(0.5949926,3.7592533){\psframe[linewidth=0.002,dimen=outer,fillstyle=gradient,gradlines=2000,gradbegin=color249g,gradend=color249f,gradmidpoint=1.0](8.4,3.3768504)(7.52,-2.0431497)}
\rput{-27.565233}(0.82673806,4.663989){\psframe[linewidth=0.002,dimen=outer,fillstyle=gradient,gradlines=2000,gradbegin=color249g,gradend=color249f,gradmidpoint=1.0](10.36,3.3568504)(9.48,-2.0631497)}
\rput{-27.565233}(1.0654684,5.5571995){\psframe[linewidth=0.002,dimen=outer,fillstyle=gradient,gradlines=2000,gradbegin=color249g,gradend=color249f,gradmidpoint=1.0](12.3,3.3168504)(11.42,-2.1031497)}
\psline[linewidth=0.04cm,linestyle=dashed,dash=0.16cm 0.16cm](0.0,2.7368505)(0.0,-0.56314963)
\psline[linewidth=0.04cm,linestyle=dashed,dash=0.16cm 0.16cm](1.72,2.7368505)(0.02,-0.62314963)
\usefont{T1}{ptm}{m}{n}
\rput(0.6326563,2.3168504){\huge $\phi$}
\end{pspicture} 
}
	\caption{The single-layer grating of Fig.\ \ref{fig1} rotated through angle $\phi$ to make a slab that is non-symmetric with respect to   the incident angle. }
	\label{fig2}
\end{figure} 

 Brewster angle total transmission is rarely observed for acoustic waves in homogeneous materials. 
The successful demonstration of EOT therefore raised  the question of whether the same    subwavelength effect can be achieved in acoustics.  \citet{D'Aguanno12} answered the question in the affirmative, demonstrating theoretically and experimentally  an acoustic grating that is completely transparent to sound waves.  The Brewster-like effect was explained via  surface impedance matching between the exterior air and the effectively rigid grating.  The gaps presented by the spaces between the grating elements allows the effective impedance to be designed to produce arbitrary Brewster angle. 
Subsequent demonstrations of extraordinary acoustic transmission (EAT) include
\citet{Qiu2012} who considered  a hybrid grating composed of two dissimilar grating elements different from the exterior air. Higher frequency properties of EAT gratings are discussed by \citet{Qi12},  while  \citet{Akbek2014} examine pass and stop-band effects for 1D phononic crystals made from repeated EAT slabs.  

The mechanism behind EAT is, as noted by \citet{D'Aguanno12}, impedance matching. The 
grating displays effective long-wavelength properties easily estimated for rigid grating elements, which allows tuning the grating porosity to achieve the desired intromission angle  \cite{D'Aguanno12}.  
\citet{Maurel13} also provide a clear explanation of the phenomenon as impedance matching but in the context of acoustics of fluids with anisotropic inertia. 
They  considered more complicated gratings comprising fluid elements and geometrical substructure, such as double layer gratings.  They showed that the anisotropic effective properties of the grating can be accurately  predicted using homogenization theory.  This opens the door to the design of EAT gratings by varying the material properties and the geometrical details.  No matter how complicated the design, homogenization theory will predict the EAT properties as long as the horizontal substructure periodicity  is subwavelength.  At shorter wavelength the present approach breaks down as dispersive effects come into play. Interesting  nonlocal effects may be expected, as has been demonstrated for  electromagnetic metamaterials comprising slanted inclusions \cite{Silveirinha2009}.  While the geometries considered here involve waveguides of  uniform width, tapered waveguides could be considered, as in \cite{Fleury2014},  introducing gradients in the effective properties. However, these possibilities are beyond the scope of this paper and remain as  future areas of study.

The purpose of this paper is to demonstrate EAT effects in non-symmetric gratings, of the type  depicted in Fig.\ \ref{fig2}.  We consider the general case in which the  grating material is an acoustic fluid, the rigid SLG being a limiting case.   We use homogenization theory to replace the grating by an equivalent effective medium with {\it anisotropic density}.  The transmission coefficient of the equivalent uniform slab can then be obtained in closed form in terms of the original  parameters, including the orientation angle of the oblique grating.  Prior to this work EAT has only been considered in the context of symmetric gratings, such as in Fig.\ \ref{fig1}.   While  \citet{Qi2015} provide experimental measurements of the effective index and effective impedance of obliquely oriented SLGs as function of wavelength, they  do not report EAT effects nor do they provide analytical results for the effective properties.  Based on the results of this paper it would be straightforward to estimate the effective index and effective impedance of SLGs as shown in Fig.\ \ref{fig2}.

The outline of the paper is as follows.  We begin in \S\ref{sec2} with a homogeneous model of an acoustic slab with anisotropic density, and derive explicit expressions for plane wave reflection and transmission, eqs.\ \eqref{64} and \eqref{65}.  The effective anisotropic properties are then derived in \S\ref{sec3} using standard long-wavelength homogenization methods.   Several limiting cases, including the rigid grating, are discussed in   \S\ref{sec4}.  Numerical examples  illustrating the dependence of intromission angle and transmittivity on the slant angle $\phi$ are given in \S\ref{sec5}.    Conclusions are presented in \S\ref{sec6}.


\section{Acoustic transmission through a slab with anisotropic inertia}\label{sec2}

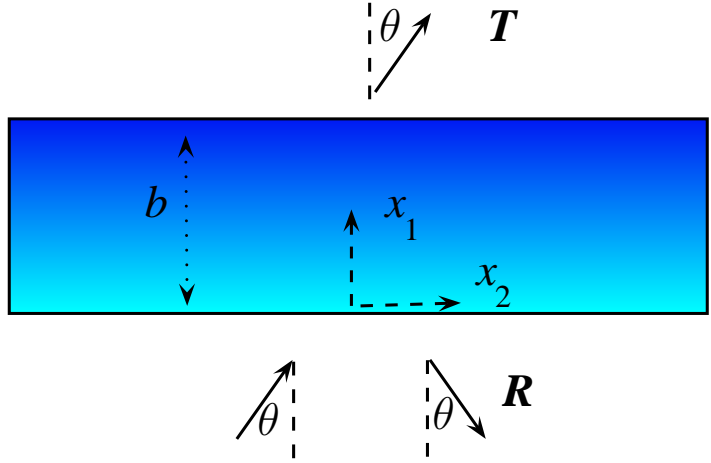
\begin{figure}[H] 
	\centering
\scalebox{1} 
{
\begin{pspicture}(0,-3.04)(9.22,3.04)
\psframe[linewidth=0.04,dimen=outer,fillstyle=gradient,gradlines=2000,gradmidpoint=1.0](9.22,1.5)(0.0,-1.12)
\psline[linewidth=0.04cm,linestyle=dashed,dash=0.16cm 0.16cm,arrowsize=0.133cm 2.0,arrowlength=1.4,arrowinset=0.4]{<-}(4.52,0.28)(4.52,-1.0)
\psline[linewidth=0.04cm,linestyle=dotted,dotsep=0.16cm,arrowsize=0.133cm 2.0,arrowlength=1.4,arrowinset=0.4]{<->}(2.34,1.26)(2.36,-1.0)
\usefont{T1}{ptm}{m}{it}
\rput(5.101875,0.3){\LARGE x}
\usefont{T1}{ptm}{m}{n}
\rput(5.3173437,0.015){\large 1}
\usefont{T1}{ptm}{m}{it}
\rput(1.95375,0.38){\LARGE b}
\psline[linewidth=0.04cm,arrowsize=0.133cm 2.0,arrowlength=1.4,arrowinset=0.4]{<-}(3.747089,-1.7157427)(3.012911,-2.7642572)
\psline[linewidth=0.04cm,arrowsize=0.133cm 2.0,arrowlength=1.4,arrowinset=0.4]{<-}(6.287089,-2.7642572)(5.5529113,-1.7157427)
\psline[linewidth=0.04cm,arrowsize=0.133cm 2.0,arrowlength=1.4,arrowinset=0.4]{<-}(5.567089,2.8842573)(4.832911,1.8357427)
\usefont{T1}{ptm}{b}{it}
\rput(6.72125,-2.12){\LARGE R}
\usefont{T1}{ptm}{b}{it}
\rput(6.5039062,2.72){\LARGE T}
\psline[linewidth=0.04cm,linestyle=dashed,dash=0.16cm 0.16cm](5.52,-1.72)(5.52,-3.0)
\usefont{T1}{ptm}{m}{it}
\rput(5.73375,-2.42){\LARGE $\theta$}
\psline[linewidth=0.04cm,linestyle=dashed,dash=0.16cm 0.16cm](3.76,-1.74)(3.76,-3.02)
\psline[linewidth=0.04cm,linestyle=dashed,dash=0.16cm 0.16cm](4.76,3.02)(4.76,1.74)
\usefont{T1}{ptm}{m}{it}
\rput(3.43375,-2.5){\LARGE $\theta$}
\usefont{T1}{ptm}{m}{it}
\rput(5.03375,2.72){\LARGE $\theta$}
\psline[linewidth=0.04cm,linestyle=dashed,dash=0.16cm 0.16cm,arrowsize=0.133cm 2.0,arrowlength=1.4,arrowinset=0.4]{<-}(5.96,-0.96)(4.64,-1.0)
\usefont{T1}{ptm}{m}{it}
\rput(6.301875,-0.6){\LARGE x}
\usefont{T1}{ptm}{m}{n}
\rput(6.51625,-0.865){\large 2}
\end{pspicture} 
}
	\caption{Two-dimensional  configuration for the equivalent uniform slab with anisotropic density.}
	\label{fig3}
\end{figure}

 The exterior acoustic medium has density $\rho$ and sound speed $c$, with bulk modulus $K=\rho c^2$. 
The governing acoustic equations for the acoustic pressure $p$ and velocity ${\bf v}$ are 
\beq{00a}
{\bf v} = (i\omega \rho)^{-1}  \nabla p ,
\ \ \
p = (i\omega )^{-1}  K \div {\bf v} .
\eeq
Time harmonic dependence $e^{-i\omega t}$ is assumed. 
The acoustic pressure comprises  incident, reflected and transmitted plane waves as shown in Fig.\ \ref{fig3},
\beq{-1}
p = p_0 \, e^{ik \sin \theta \, x_2  }  \times\begin{cases}
\big( e^{i k\cos \theta\,  x_1} + R e^{-i k \cos \theta\,  x_1}\big) & x_1 \le 0,
\\  
Te^{i k \cos \theta  (x_1 -\a)}   & x_1 \ge \a,
\end{cases}
\eeq
where $k = \frac{\omega}c$ and $p_0$ is a constant.   
Define for later use the acoustic impedance
\beq{-32}
Z_\theta = \frac{ \rho c}{\cos\theta} . 
\eeq
 We are interested in conditions for which $|T|=1$.  

Consider a uniform slab of thickness $b$,  bulk modulus $K_s$ and   inertia tensor which is represented by a 
 2$\times$2  symmetric matrix 
$(\boldsymbol{\rho} =\boldsymbol{\rho}^T)$ with elements $\rho_{ij}$, $i,j=1,2$.   Specific models for anisotropic non-diagonal  density tensors are discussed in \S\ref{sec3}.  The equations of motion within the slab are
\beq{006}
{\bf v} = (i\omega \boldsymbol{\rho})^{-1}  \nabla p ,
\ \ \
p = (i\omega )^{-1}  K_s \div {\bf v} .
\eeq
The transmission and reflection coefficients follow from Appendix \ref{321} as 
\bse{64}
\bal{64a}
T &= e^{-ik \a \frac{\rho_{12}}{\rho_{22}}\sin\theta }
\Big( \cos \frac{\omega \a}{c_\theta} - \frac i2 \Big( \frac{Z_\theta}{Z_\theta'} 
+\frac{Z_\theta'}{Z_\theta}  \Big) 
\sin  \frac{\omega \a}{c_\theta}  \Big)^{-1},
\\
R &= \frac i2 \Big(  \frac{Z_\theta}{Z_\theta'}  - \frac{Z_\theta'}{Z_\theta}  \Big)
\sin  \frac{\omega \a}{c_\theta} 
\Big( \cos \frac{\omega \a}{c_\theta} - \frac i2 
\Big( \frac{Z_\theta}{Z_\theta'} 
+\frac{Z_\theta'}{Z_\theta}  \Big) 
\sin  \frac{\omega \a}{c_\theta}  \Big)^{-1},
\label{64b}
\eal
\ese
where   
\beq{65}
 c_\theta =\Big(   \frac{\rho_{22}} {\det \boldsymbol{\rho}}    \Big)^{\frac 12}
\Big(  \frac 1{K_s} - \frac{\sin^2\theta}{c^2\rho_{22} }  \Big)^{-\frac 12} ,
\quad 
 Z_\theta' =  
\Big(   \frac{\det \boldsymbol{\rho}}{\rho_{22}} \Big)c_\theta .
\eeq
The results \eqref{64}-\eqref{65} have not been previously published.  Maurel et al. \cite{Maurel13} consider the particular case of aligned inertial and slab axes $(\rho_{12} =0)$.  
The case of a normal acoustic fluid in the slab corresponds to $\rho_{11}=\rho_{22}$, 
$\rho_{12} =0$. 
Note that 
\beq{6-56}
R(-\theta)=R(\theta) , \quad |T(-\theta)| =|T(\theta)|
\eeq
Thus, as a function of incident angle the reflection coefficient is  symmetric about $\theta = 0$, but only the magnitude of the transmission coefficient is symmetric.  The asymmetry as a function of $\theta $ is evident from 
the relation
\beq{6-6}
\frac{T(-\theta) }{T(\theta) } = e^{i2 k \a \frac{\rho_{12}}{\rho_{22}}\sin\theta }.
\eeq
A similar expression for  the ratio of transmission coefficients 
for transmission through an anisotropic dielectric slab was derived by Castani\'e et al.\  
\cite[Eqs. (18) to (21)]{Castanie2014}, who also considered propagation through layered anisotropic dielectric media.  Note that the identity \eqref{6-56}$_1$ for the reflection coefficient 
is expected based on  reciprocity \cite{Achenbach04}.

Equation \eqref{64a} implies $|T|=1$ when the {\it impedances match},  
\beq{66}
|T(\theta_i)|=1 \ \ \Leftrightarrow \ \  Z_\theta = Z_\theta' . 
\eeq
Hence, enhanced acoustic transmittivity occurs if 
\beq{646}
\rho^2 \sin^2 \theta_i + (\det \boldsymbol{\rho} ) \cos^2 \theta_i 
= \frac{K}{K_s}\rho \rho_{22} . 
\eeq
The value of  $\theta_i$ satisfying this relation is the { intromission angle}.  It is clear from $\eqref{6-56}_2$ that the intromission effect is symmetric in the incident angle, i.e.\ 
$|T(\pm \theta_i)|=1$.

\section{Single-layer gratings as anisotropic inertial slabs}\label{sec3}

\begin{figure}[h] 
	\centering
{\parbox[c][1.5in][t]{0.45\textwidth}{
\scalebox{.45} 
{
\begin{pspicture}(0,-3.7965624)(14.055,3.7965624)
\definecolor{color245g}{rgb}{0.8,0.0,1.0}
\definecolor{color245f}{rgb}{0.8,0.0,0.8}
\psframe[linewidth=0.002,dimen=outer,fillstyle=gradient,gradlines=2000,gradbegin=color245g,gradend=color245f,gradmidpoint=1.0](2.095,3.113125)(1.215,-2.306875)
\psframe[linewidth=0.002,dimen=outer,fillstyle=gradient,gradlines=2000,gradbegin=color245g,gradend=color245f,gradmidpoint=1.0](3.795,3.093125)(2.915,-2.326875)
\psframe[linewidth=0.002,dimen=outer,fillstyle=gradient,gradlines=2000,gradbegin=color245g,gradend=color245f,gradmidpoint=1.0](5.515,3.073125)(4.635,-2.346875)
\psframe[linewidth=0.002,dimen=outer,fillstyle=gradient,gradlines=2000,gradbegin=color245g,gradend=color245f,gradmidpoint=1.0](7.215,3.053125)(6.335,-2.366875)
\psframe[linewidth=0.002,dimen=outer,fillstyle=gradient,gradlines=2000,gradbegin=color245g,gradend=color245f,gradmidpoint=1.0](8.935,3.073125)(8.055,-2.346875)
\psframe[linewidth=0.002,dimen=outer,fillstyle=gradient,gradlines=2000,gradbegin=color245g,gradend=color245f,gradmidpoint=1.0](10.635,3.053125)(9.755,-2.366875)
\psframe[linewidth=0.002,dimen=outer,fillstyle=gradient,gradlines=2000,gradbegin=color245g,gradend=color245f,gradmidpoint=1.0](12.355,3.033125)(11.475,-2.386875)
\psframe[linewidth=0.002,dimen=outer,fillstyle=gradient,gradlines=2000,gradbegin=color245g,gradend=color245f,gradmidpoint=1.0](14.055,3.013125)(13.175,-2.406875)
\psline[linewidth=0.04cm,arrowsize=0.05291667cm 8.0,arrowlength=1.4,arrowinset=0.4]{<->}(0.435,3.093125)(0.475,-2.206875)
\usefont{T1}{ptm}{m}{n}
\rput(0.1171875,0.718125){\huge a}
\psline[linewidth=0.04cm,arrowsize=0.05291667cm 4.0,arrowlength=1.4,arrowinset=0.4]{<->}(2.935,-3.046875)(1.215,-3.046875)
\psline[linewidth=0.04cm,linestyle=dashed,dotsep=0.16cm](1.215,-2.286875)(1.215,-3.046875)
\psline[linewidth=0.04cm,linestyle=dashed,dotsep=0.16cm](2.955,-2.406875)(2.955,-3.166875)
\usefont{T1}{ptm}{m}{n}
\rput(2.0785937,-3.501875){\huge d}
\psline[linewidth=0.04cm,linestyle=dashed,dotsep=0.16cm](4.635,-2.406875)(4.635,-3.166875)
\psline[linewidth=0.04cm,linestyle=dashed,dotsep=0.16cm](5.495,-2.406875)(5.495,-3.166875)
\psline[linewidth=0.04cm,arrowsize=0.05291667cm 4.0,arrowlength=1.4,arrowinset=0.4]{<->}(5.515,-3.066875)(4.615,-3.066875)
\usefont{T1}{ptm}{m}{n}
\rput(5.054375,-3.541875){\huge fd}
\usefont{T1}{ptm}{m}{n}
\rput(3.1648438,3.498125){\Large $K_0$, $\rho_0$}
\end{pspicture} 
}} } 
~~~
{\parbox[c][1.5in][s]{0.45\textwidth}{
\scalebox{.45} 
{
\begin{pspicture}(0,-3.5328372)(14.1,3.492488)
\definecolor{color249g}{rgb}{0.8,0.0,1.0}
\definecolor{color249f}{rgb}{0.8,0.0,0.8}
\rput{-27.565233}(-0.16747421,1.0909739){\psframe[linewidth=0.002,dimen=outer,fillstyle=gradient,gradlines=2000,gradbegin=color249g,gradend=color249f,gradmidpoint=1.0](2.58,3.5968504)(1.7,-1.8231496)}
\psline[linewidth=0.04cm,arrowsize=0.05291667cm 8.0,arrowlength=1.4,arrowinset=0.4]{<->}(14.08,2.6368504)(11.68,-2.0431497)
\usefont{T1}{ptm}{m}{n}
\rput(13.362187,0.3018504){\huge a}
\psline[linewidth=0.04cm,arrowsize=0.05291667cm 4.0,arrowlength=1.4,arrowinset=0.4]{<->}(3.7,-2.9231496)(2.16,-2.1831496)
\psline[linewidth=0.04cm,linestyle=dashed,dotsep=0.16cm](2.42,-1.4831496)(2.06,-2.1831496)
\usefont{T1}{ptm}{m}{n}
\rput(2.6435938,-2.9981496){\huge d}
\psline[linewidth=0.04cm,arrowsize=0.05291667cm 4.0,arrowlength=1.4,arrowinset=0.4]{<->}(6.5,-3.0031495)(5.78,-2.5831497)
\usefont{T1}{ptm}{m}{n}
\rput(5.859375,-3.2781496){\huge fd}
\usefont{T1}{ptm}{m}{n}
\rput(10.049844,-2.5){\huge $K_0$, $\rho_0$}
\rput{-27.565233}(0.07125611,1.9841841){\psframe[linewidth=0.002,dimen=outer,fillstyle=gradient,gradlines=2000,gradbegin=color249g,gradend=color249f,gradmidpoint=1.0](4.52,3.5568504)(3.64,-1.8631496)}
\rput{-27.565233}(0.35626227,2.8660429){\psframe[linewidth=0.002,dimen=outer,fillstyle=gradient,gradlines=2000,gradbegin=color249g,gradend=color249f,gradmidpoint=1.0](6.46,3.4168503)(5.58,-2.0031495)}
\rput{-27.565233}(0.5949926,3.7592533){\psframe[linewidth=0.002,dimen=outer,fillstyle=gradient,gradlines=2000,gradbegin=color249g,gradend=color249f,gradmidpoint=1.0](8.4,3.3768504)(7.52,-2.0431497)}
\rput{-27.565233}(0.82673806,4.663989){\psframe[linewidth=0.002,dimen=outer,fillstyle=gradient,gradlines=2000,gradbegin=color249g,gradend=color249f,gradmidpoint=1.0](10.36,3.3568504)(9.48,-2.0631497)}
\rput{-27.565233}(1.0654684,5.5571995){\psframe[linewidth=0.002,dimen=outer,fillstyle=gradient,gradlines=2000,gradbegin=color249g,gradend=color249f,gradmidpoint=1.0](12.3,3.3168504)(11.42,-2.1031497)}
\psline[linewidth=0.04cm,linestyle=dashed,dotsep=0.16cm](4.4,-1.5831496)(3.72,-2.8231497)
\psline[linewidth=0.04cm,linestyle=dashed,dotsep=0.16cm](6.3,-1.6831496)(5.76,-2.6431496)
\psline[linewidth=0.04cm,linestyle=dashed,dotsep=0.16cm](6.98,-2.0031495)(6.46,-2.9231496)
\psline[linewidth=0.04cm,linestyle=dashed,dash=0.16cm 0.16cm](0.0,2.7368505)(0.0,-0.56314963)
\psline[linewidth=0.04cm,linestyle=dashed,dash=0.16cm 0.16cm](1.72,2.7368505)(0.02,-0.62314963)
\usefont{T1}{ptm}{m}{n}
\rput(0.6326563,2.3168504){\huge $\phi$}
\end{pspicture} 
}} }
	\caption{A single-layer grating.  The grating material is an acoustic fluid of bulk modulus  $K_0$, density $\rho_0$ and  volume fraction  $f$.   A symmetric  gratings is shown on the left.  On the  right,    
	the elements of the  grating are rotated through angle $\phi$ to make a slab that is non-symmetric with respect to   the incident angle $\theta$ of Fig.\ \ref{fig3}. 	
	}
	\label{fig4}
\end{figure}
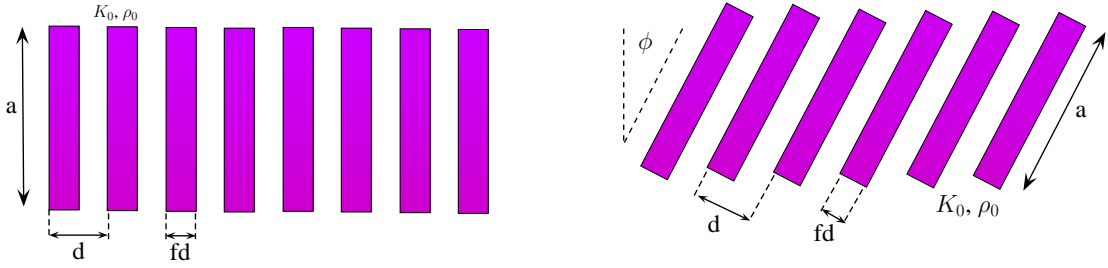 

\begin{figure}[h] 
	\centering
\includegraphics[width=0.5\textwidth]{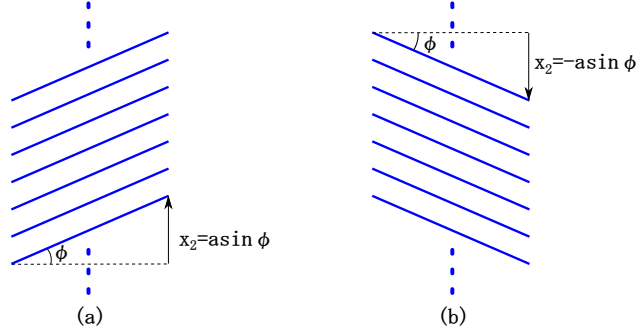}
	\caption{Relative phase of the transmission coefficients for $\pm\theta$.}
	\label{phase}
\end{figure}

Consider first the symmetric  single-layer grating (SLG) of Fig.\ \ref{fig4} 
for which  the grating fluid has  properties $K_0$, $\rho_0$, and  volume fraction $f\in[0,1]$. 
The  effective bulk modulus $K_s$ and density tensor $\boldsymbol{\rho}$ of the slab follow from standard quasi-static homogenization, e.g.\  \cite[eq.\ (3)]{Maurel13}, as 
\beq{92}
\frac 1{K_s}  = \frac f{K_0} + \frac{1-f}K, \ \
\boldsymbol{\rho} = 
\begin{pmatrix} \rho_1 & 0 \\ 0 & \rho_2
\end{pmatrix}
, \ \
\frac 1{\rho_1} = \frac{f}{\rho_0} + \frac{1-f}{\rho} , 
\ \
 \rho_2= f\rho_0 + (1-f)\rho .
\eeq
The SLG rotated through angle $\phi$ relative to the $x_1x_2$ directions as in  Fig.\ \ref{fig4}  has the same effective bulk modulus $K_s$ while the inertia tensor becomes non-diagonal and  symmetric,   with 
\beq{93}
\boldsymbol{\rho} = 
\begin{pmatrix} \rho_{11} & \rho_{12} \\ \rho_{21} & \rho_{22}
\end{pmatrix}, \ \ \ 
\begin{aligned}
\rho_{11} & = \rho_1 \cos^2\phi + \rho_2 \sin^2\phi ,
\\
\rho_{22} & = \rho_1 \sin^2\phi + \rho_2 \cos^2\phi ,
\\
\rho_{12} & = (\rho_1- \rho_2) \sin \phi\cos\phi \ \big(=\rho_{21}\big). 
\end{aligned}
\eeq
Note that  $\rho_{11} , \rho_{22} >0$,
\beq{94}
\rho_{12}  = - \, \frac{f(1-f) (\rho- \rho_0)^2}{f\rho + (1-f)\rho_0} \sin \phi\cos\phi  
\ \ \Rightarrow \ \
\begin{cases}
\rho_{12} <0 & \text{if} \ \phi >0,
\\
\rho_{12} >0 & \text{if} \ \phi <0,
\\
\rho_{12} =0 & \text{if} \ \phi =0,
\end{cases}
\eeq
while $\det \boldsymbol{\rho} =\rho_1 \rho_2$ and $\tr \boldsymbol{\rho} =\rho_1 + \rho_2$ are independent of $\phi$.  
The relative phase of the transmitted wave for incidence at $\pm \theta$, eq.\ \eqref{6-6}, becomes, using $\a=a \cos\phi$, 
\beq{6-612}
\frac{T(-\theta) }{T(\theta) } = e^{-i2 k a \sin\phi\sin\theta \, \big(\frac{ \rho_2-\rho_1}{\rho_2+\rho_1 \tan^2\phi}\big)
}.
\eeq
The phase difference in \eqref{6-612} between $T(\theta)$ and $T(-\theta)$ can be understood as follows.  First, the term $-2 k a \sin\phi\sin\theta$ has clear geometrical meaning.  Referring to  Figure \ref{phase}, note that $k\sin\theta$ is the conserved horizontal wavenumber, while $a\sin\phi$ is the horizontal path length, resulting in the phase advance/delay of $\pm k a \sin\phi\sin\theta$ for incidence at $\pm  \theta$.
The additional factor, $\frac{ \rho_2-\rho_1}{\rho_2+\rho_1 \tan^2\phi}$, which is positive but less than unity on account of the fact that $\rho_2 > \rho_1$, arises from   acoustic propagation in the grating elements.  This results in a smaller phase effect than that of the rigid limit $(\rho_2 \gg \rho_1)$.
The phase term obviously becomes zero in the symmetric limit $\phi =0$. Note, however, that  for  the fluid grating the phase also tends to zero as $\phi \to \frac{\pi}2$.  

Using the explicit formulae for the  anisotropic density tensor, the impedance matching condition \eqref{646} becomes
\beq{67}
|T(\theta_i)|=1 \ \ \Leftrightarrow \ \
\boxed{\cos^2 \theta_i 
= \frac{ \big(   \frac{\rho_2-\rho_1}{\rho}\cos^2\phi  +\frac{\rho_1}{\rho} \big)\frac K{K_s} -1} {\frac{ \rho_1\rho_2}{\rho^2} -1} } 
\eeq
This shows the explicit dependence on the orientation angle $\phi$.  In particular, it 
 implies that $\partial \theta_i /\partial \phi >0$ since  
$\rho_2-\rho_1 > 0$ and $\rho_1\rho_2 > \rho^2$ for $\rho_0\ne \rho$.

\section{Limiting cases and generalizations}\label{sec4}

The intromission angle for the single-layer grating of Fig.\ \ref{fig3} is given by eq.\ \eqref{67}. Here we consider its behavior for some limits of the parameters, such as rigid grating elements.  The limiting case of $\phi = 0$ was considered by \cite{Maurel13}, although they do not provide a simple full-transmission condition  analogous to \eqref{67} with $\phi = 0$. 

\subsection{Rigid grating elements}
If the grating element is much stiffer than the background fluid, then in the limit $K/K_0 \to 0$ \eqref{67} becomes
 \beq{67-1}
|T(\theta_i)|=1, \  \frac{K}{K_0}=0,  \ \ \Leftrightarrow \ \
{\cos^2 \theta_i 
= \frac{ (1-f)\frac{ \rho_{22}}{\rho}  -1} {\frac{ \rho_1\rho_2}{\rho^2} -1} }.
\eeq
The case  of a fixed rigid grating is obtained in the dual limit of large stiffness and density, i.e.  
 \beq{67-2}
|T(\theta_i)|=1, \  \frac{K}{K_0}=0,  \  \frac{\rho}{\rho_0}=0,  \ \ \Leftrightarrow \ \
\boxed{\cos \theta_i 
= (1-f) \cos\phi }
\eeq
 This case, which we call  the rigid limit, is of particular interest.  It is easily realized if the background acoustic medium is air. 
 
In the rigid  limit \eqref{67-2} we have (see eq.\ \eqref{65}) $c_\theta = c \cos\phi$, 
and  the transmission coefficient of \eqref{64a} simplifies to   
\beq{6-60}
T(\theta) = e^{i k a\sin \phi \sin\theta } /
\Big( \cos k a - \frac i2 \big( \frac{\cos\theta}{\cos\theta_i}+\frac{\cos\theta_i}{\cos\theta}\big) 
\sin k a\Big) .
\eeq
Hence, 
\beq{6-62}
|T(\theta)| = \cos \gamma, \ \ 
|R(\theta)| = \sin \gamma, \ \ \gamma=
\tan^{-1} \Big( \frac 12 \big( \frac{\cos\theta}{\cos\theta_i}-\frac{\cos\theta_i}{\cos\theta}\big) 
\sin   ka \Big)
\eeq
and 
the relative phase of the transmission coefficients for $\pm \theta$ is
\beq{6-61}
\frac{T(-\theta) }{T(\theta) } = e^{-i2 k a\sin\phi \sin\theta }.
\eeq
The reason for the phase difference $-2 k a\sin\phi \sin\theta $ is evident from Fig.\ \ref{phase}.  

\subsection{Transmission at normal incidence: $\theta_i = 0$ }  The  intromission angle is identically zero if 
\beq{72}
|T(0)|=1  \ \ \Leftrightarrow \ \  
\boxed{
\frac{K}{K_0  }  = \frac{
\rho_0^2 - (1-f)^2
( \rho_0- \rho)^2 \cos^2\phi  }
{\rho\rho_0 + f(1-f)(\rho_0 -\rho)^2 \cos^2\phi}
 \le \frac{\rho_0}\rho
}
\eeq
This is a rather   interesting identity: it indicates that the required impedance
ratio $\sqrt{K\rho/ K_0\rho_0}$  depends on the density ratio and the "`environmental" parameters 
$f$ and $\phi$ but not on the relative bulk moduli.  If any one of the three conditions $f=1$, $\phi = \frac{\pi}2$ or $\rho_0 = \rho$ holds then  \eqref{72} reduces to the expected one-dimensional impedance matching condition $K_0 \rho_0 = K\rho $.  However, when 
$f\ne 1$ and  $\rho_0 \ne  \rho$ eq.\ \eqref{72}  implies that 
the grating material must have higher impedance than the background fluid.

Assume further that full transmission at $\theta_i = 0$ corresponds to $\phi = 0$, then \eqref{67} (or \eqref{72}) requires 
\beq{74}
\left.
\begin{aligned}
|T(0)|&=1,\\ \phi &=0, \end{aligned}\right\}
\ \ \Leftrightarrow \ \
\frac{K}{K_0  }  = \frac{
\rho_0^2 - (1-f)^2
( \rho_0- \rho)^2   }
{\rho\rho_0 + f(1-f)(\rho_0 -\rho)^2 }
.
\eeq
Now vary  $\phi$, with \eqref{74} satisfied, then eq.\ \eqref{67} becomes 
\beq{75}
|T(\theta_i)|=1 \  \ \Leftrightarrow \ \
\sin\theta_i = \sqrt{\frac{(1-f)\rho_0}{f\rho + (1-f)\rho_0}
\Big(\frac{\rho_0-\rho}{\rho_0+\rho}\Big) }\, 
\sin\phi . 
\eeq
This provides a possible {\it active} model for changing the angular receptivity of the slab by rotating the elements of the single-layer grating.    

\subsection{Zigzag structures}
\begin{figure}[h!] 
	\centering
\includegraphics[width=0.6\textwidth]{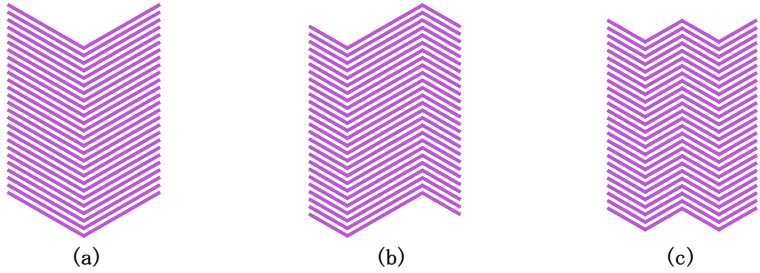}
	\caption{Zigzag structures are SLGs in series with alternating orientations $\pm \phi$.} 
	\label{zigzag}
\end{figure}

A zigzag structure, as shown in Fig. \ref{zigzag}, here means one that is made from layers in series, each layer being a SLG with grating elements oriented at $ \phi$ or $-\phi$.  The only difference between adjacent layers is that the effective density $\rho_{12}$ changes sign.   Let $b_+$ $(b_-)$ be the combined thickness of the layers with orientation $ +\phi$ $(-\phi)$, so that the total thickness is $b=b_+ +b_-$.
The transmission coefficient is 
\beq{306}
T(\theta) = e^{-ik (b_+-b_-) \frac{\rho_{12}}{\rho_{22}}\sin\theta }
\Big( \cos \frac{\omega \a}{c_\theta} - \frac i2 \Big( \frac{Z_\theta}{Z_\theta'} 
+\frac{Z_\theta'}{Z_\theta}  \Big) 
\sin  \frac{\omega \a}{c_\theta}  \Big)^{-1},
\eeq
 the reflection coefficient is given by eq.\ \eqref{64b}, and the other parameters in \eqref{306} are as before.  
The three examples of zigzag structures in Fig. \ref{zigzag} all have  $b_+=b_-$ and therefore $T(-\theta) =T(\theta)$ in each case.

\section{Numerical examples}\label{sec5}

The examples presented use non-dimensional parameters as far as possible; in particular  the frequency is defined by $kd$.   The length of the grating elements is   $a = 20d$, see Fig.\ \ref{fig4}, and the total slab thickness $\a$ depends on the orientation angle through $\a = a\cos\phi$.    All results shown were generated with  COMSOL using periodic boundary conditions to simulate wave transmission through an infinitely periodic structure.

\begin{figure}[H] 
	\centering
\renewcommand{\thesubfigure}{(a)}\subfigure[$f=0.5$, $kd=0.25$]{
\includegraphics[width=0.34\textwidth]{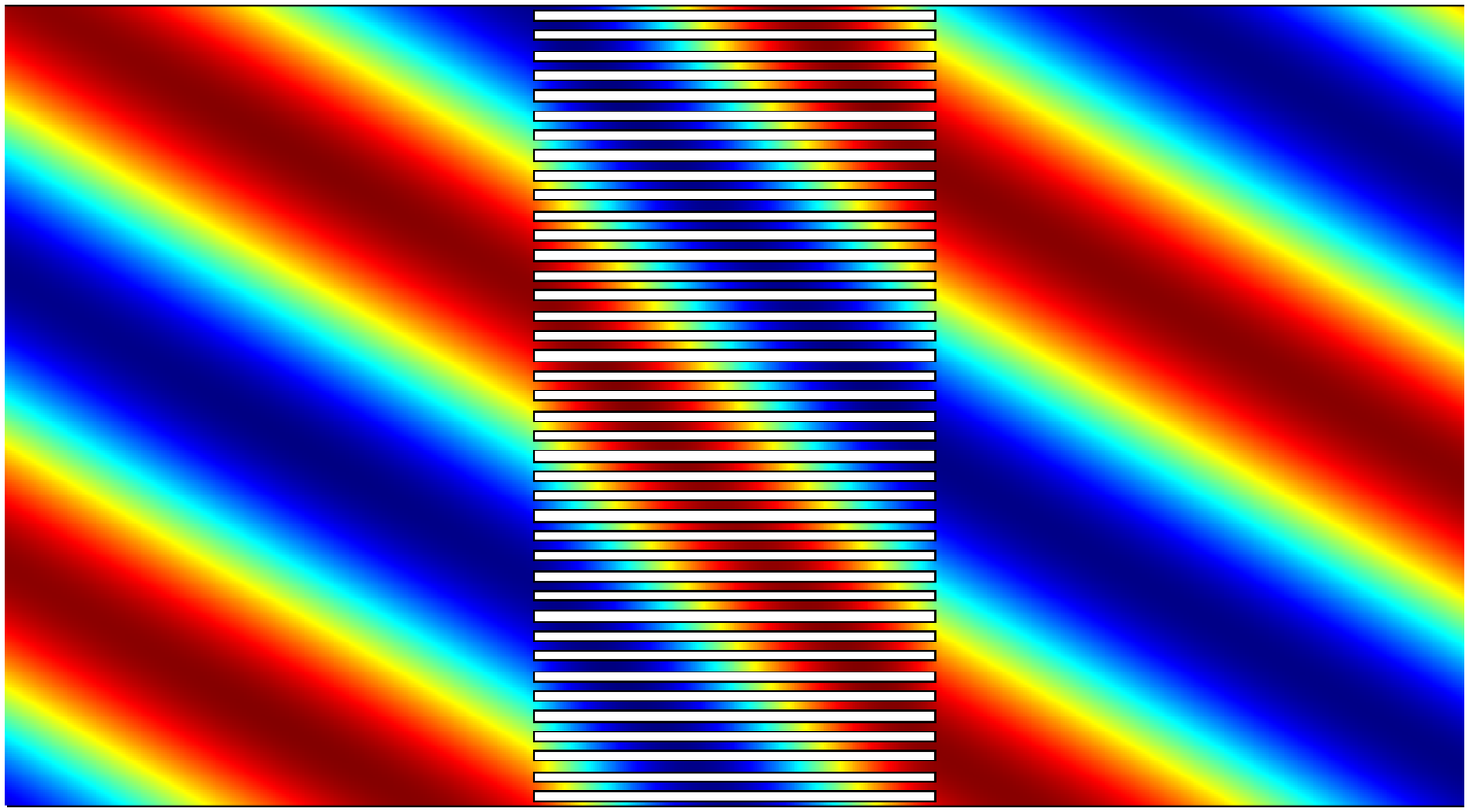} \label{pm0kd025}}
\renewcommand{\thesubfigure}{(e)}\subfigure[$f=0.75$, $kd=0.25$]{
\includegraphics[width=0.34\textwidth]{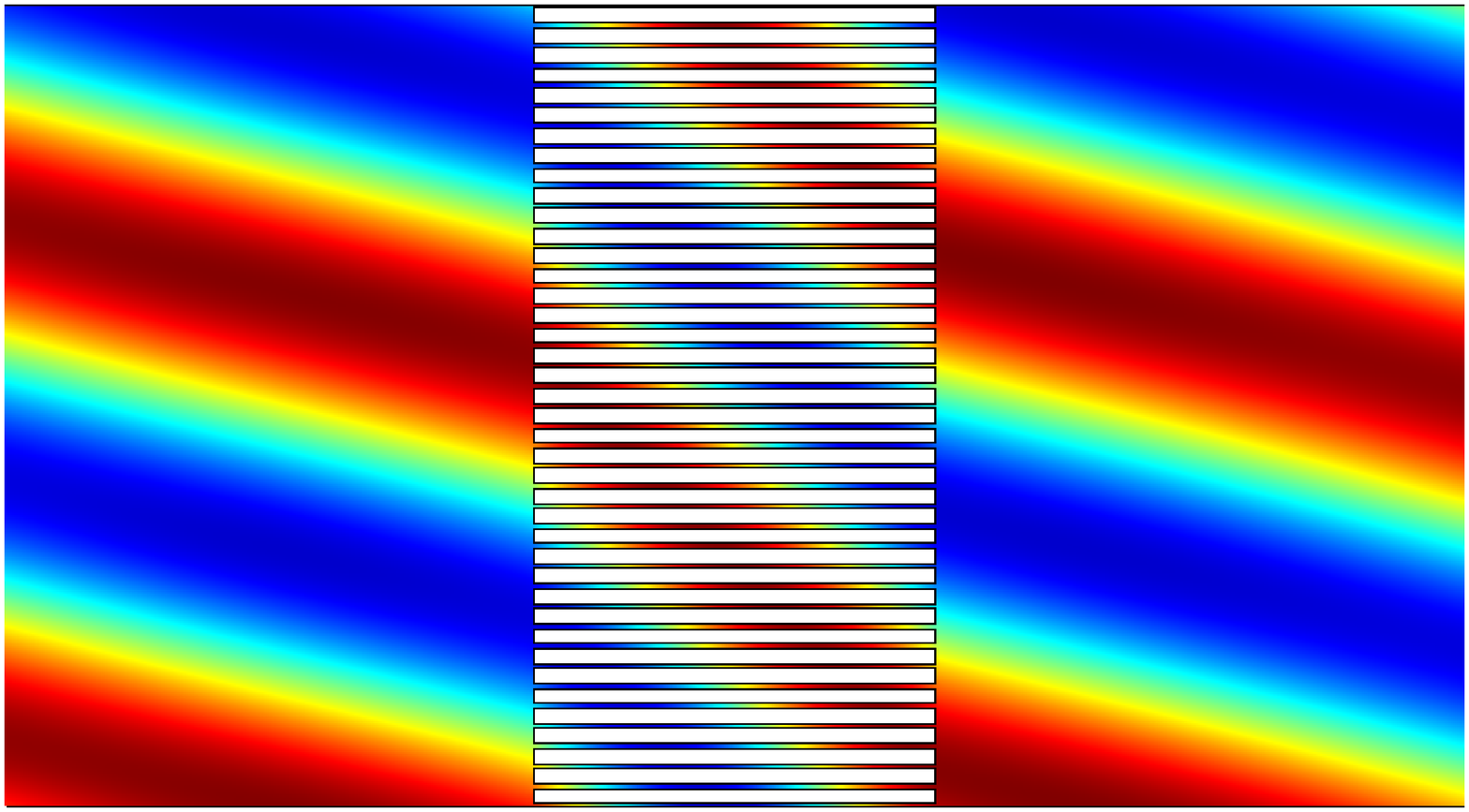} \label{pm0kd05}}\\
\renewcommand{\thesubfigure}{(b)}\subfigure[$f=0.5$, $kd=0.5$]{
\includegraphics[width=0.34\textwidth]{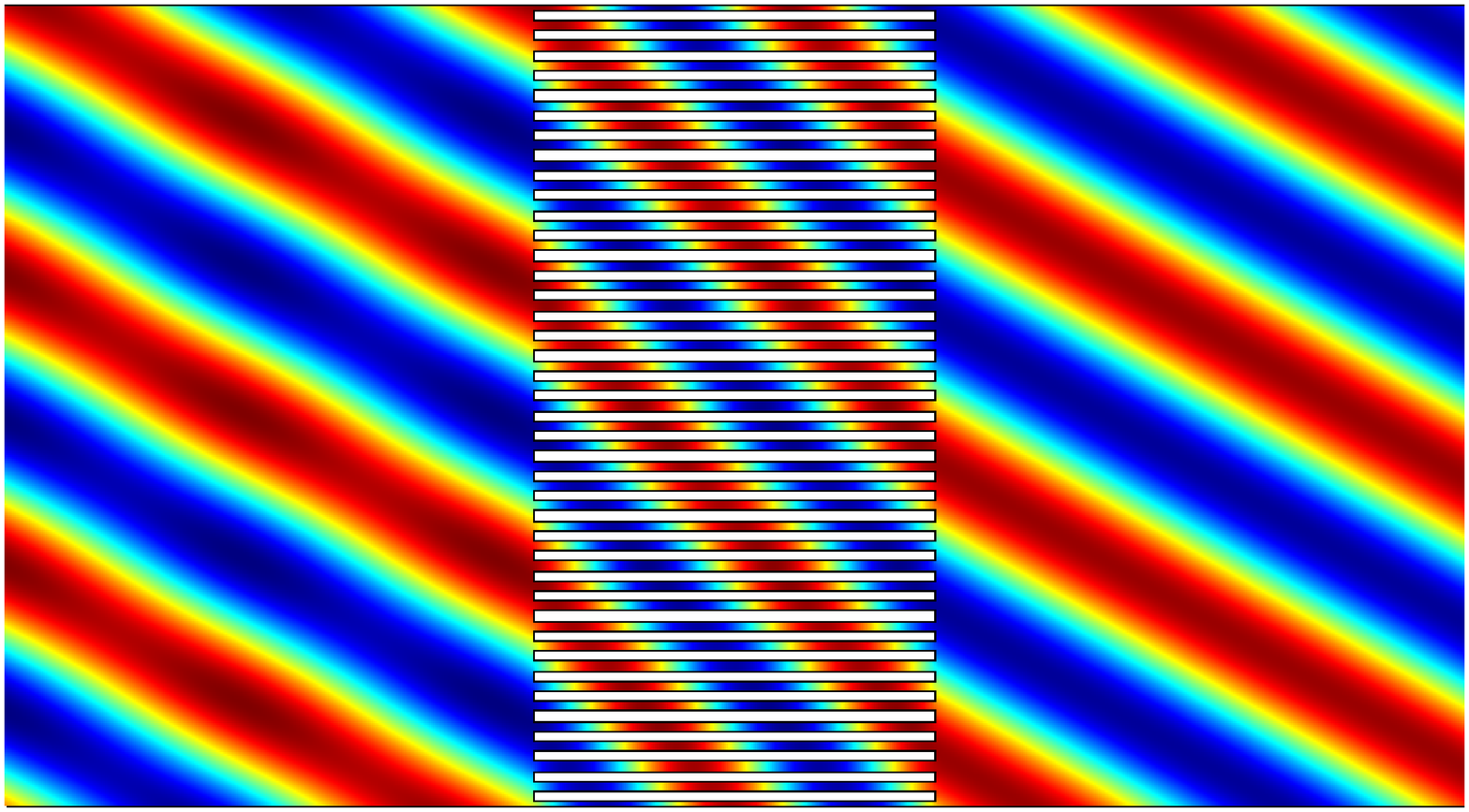} \label{pm0kd075}}
\renewcommand{\thesubfigure}{(f)}\subfigure[$f=0.75$, $kd=0.5$]{
\includegraphics[width=0.34\textwidth]{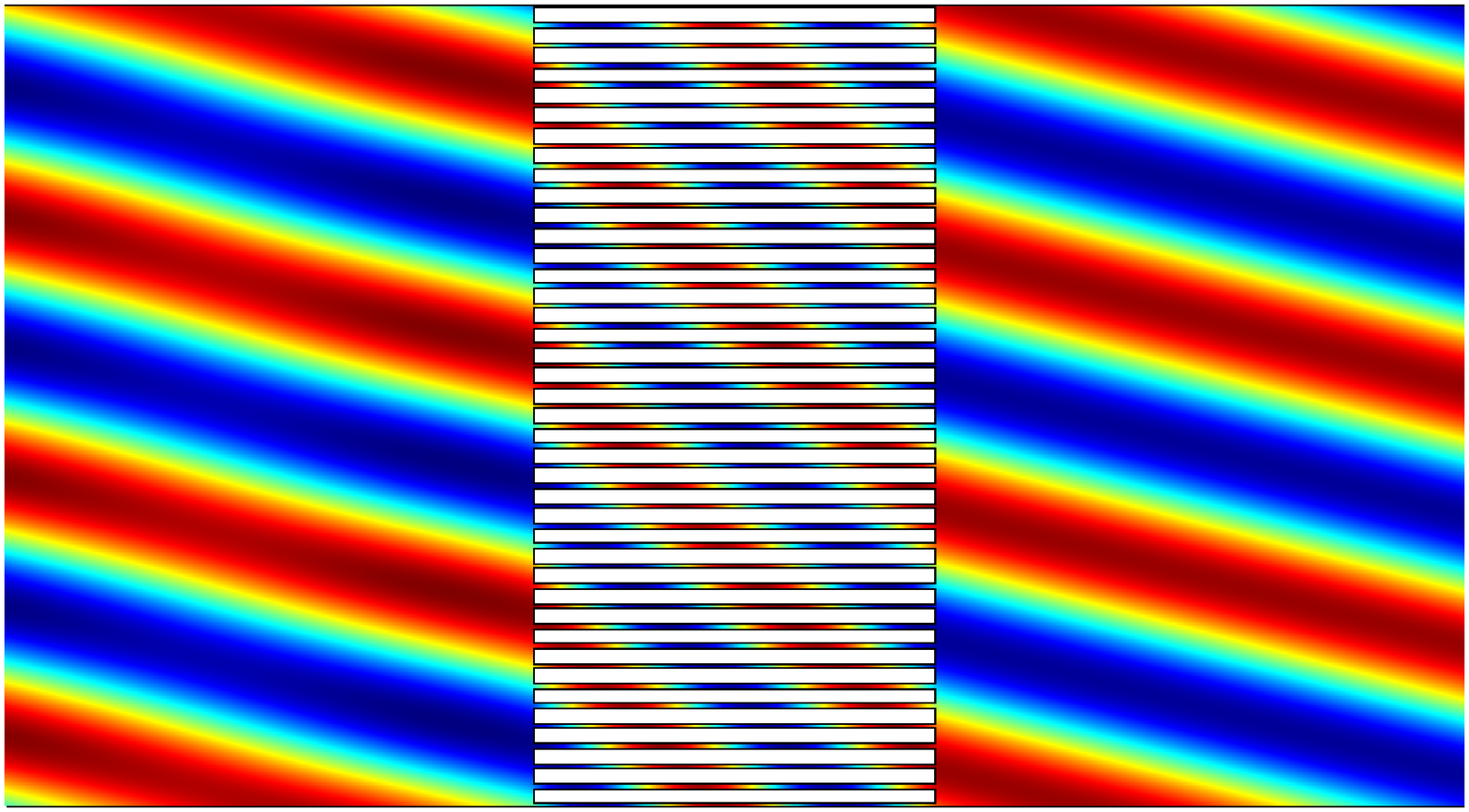} \label{pm0kd1}}\\
\renewcommand{\thesubfigure}{(c)}\subfigure[$f=0.5$, $kd=0.75$]{
\includegraphics[width=0.34\textwidth]{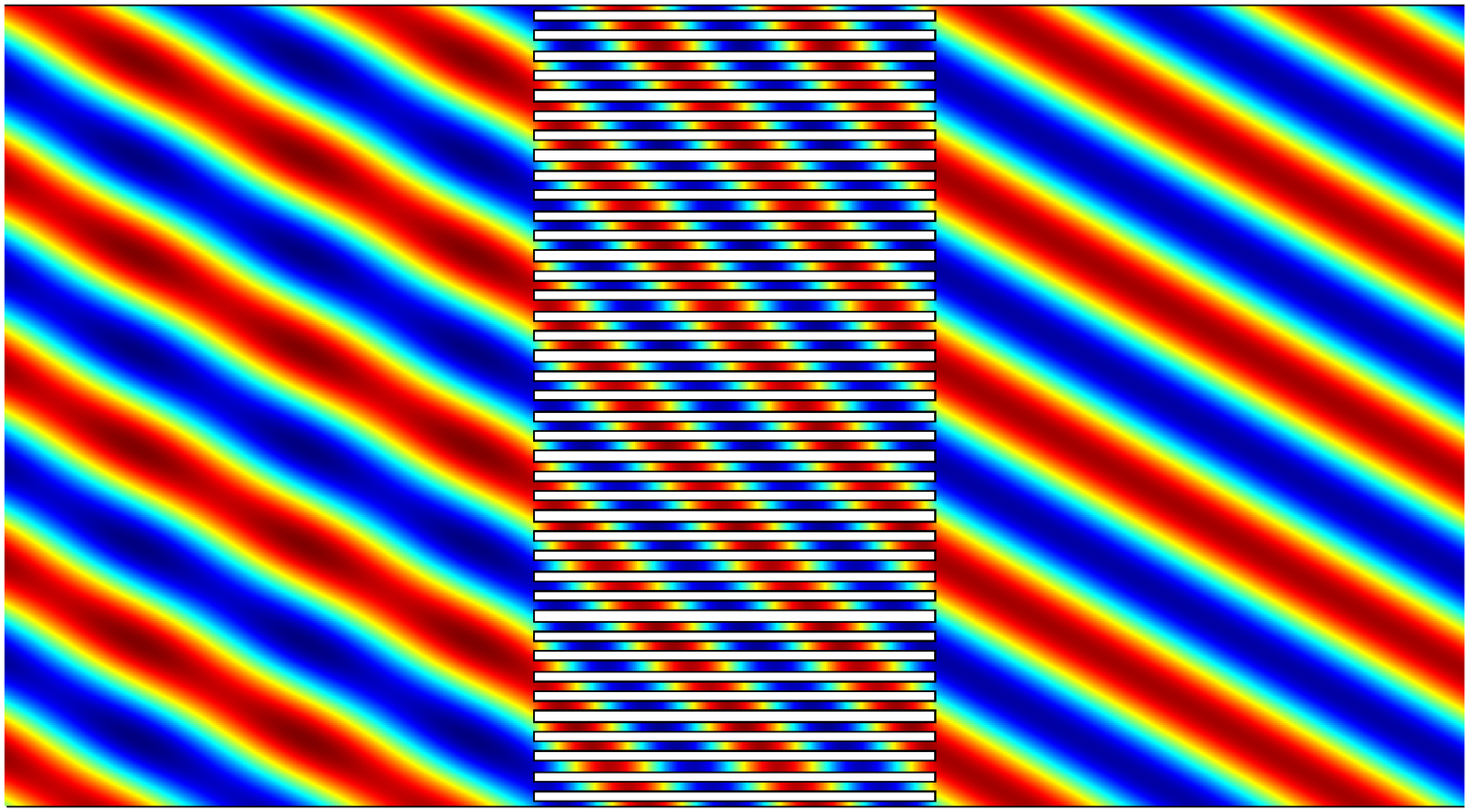} \label{pm0kd0252}}
\renewcommand{\thesubfigure}{(g)}\subfigure[$f=0.75$, $kd=0.75$]{
\includegraphics[width=0.34\textwidth]{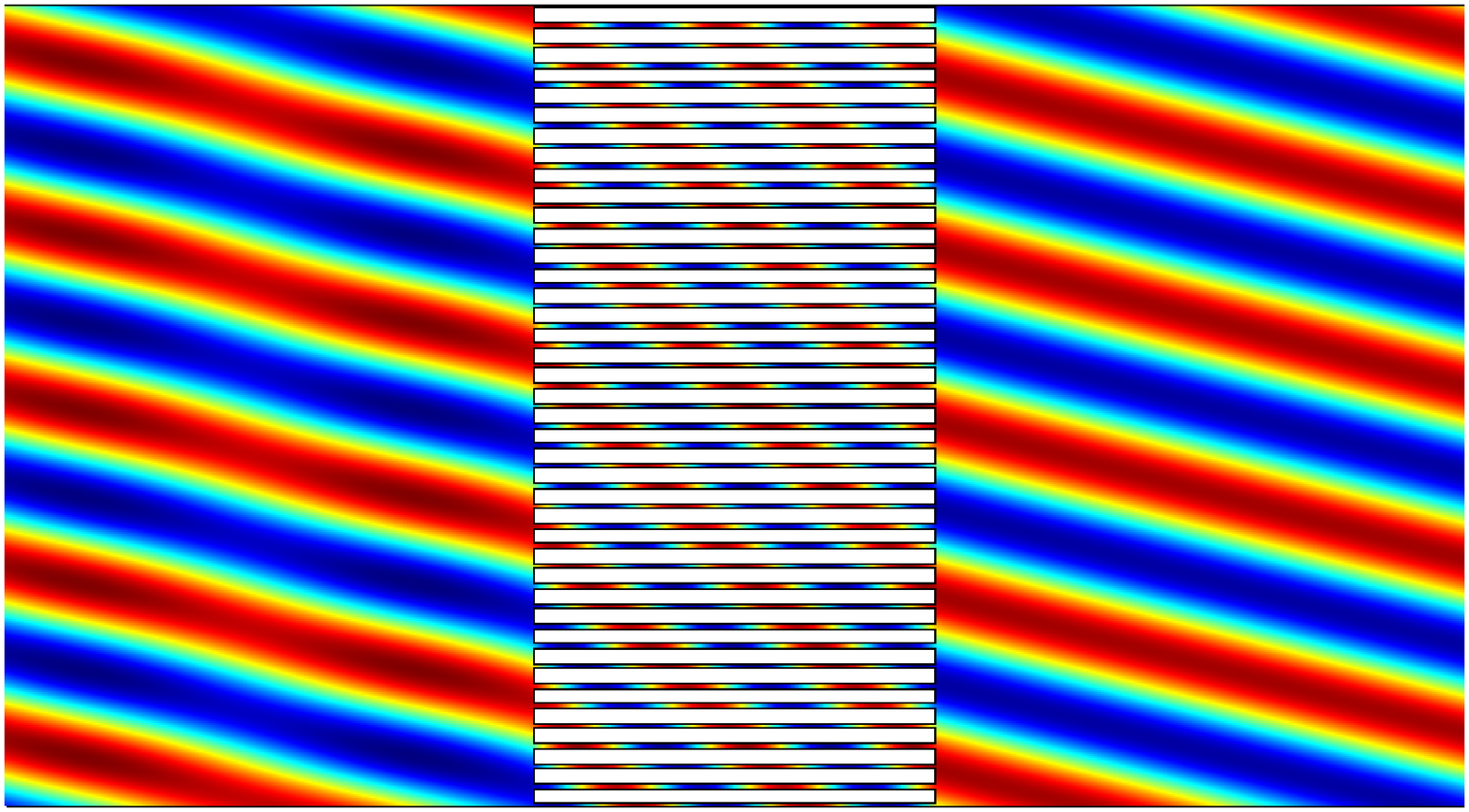} \label{pm0kd052}}\\
\renewcommand{\thesubfigure}{(d)}\subfigure[$f=0.5$, $kd=1$]{
\includegraphics[width=0.34\textwidth]{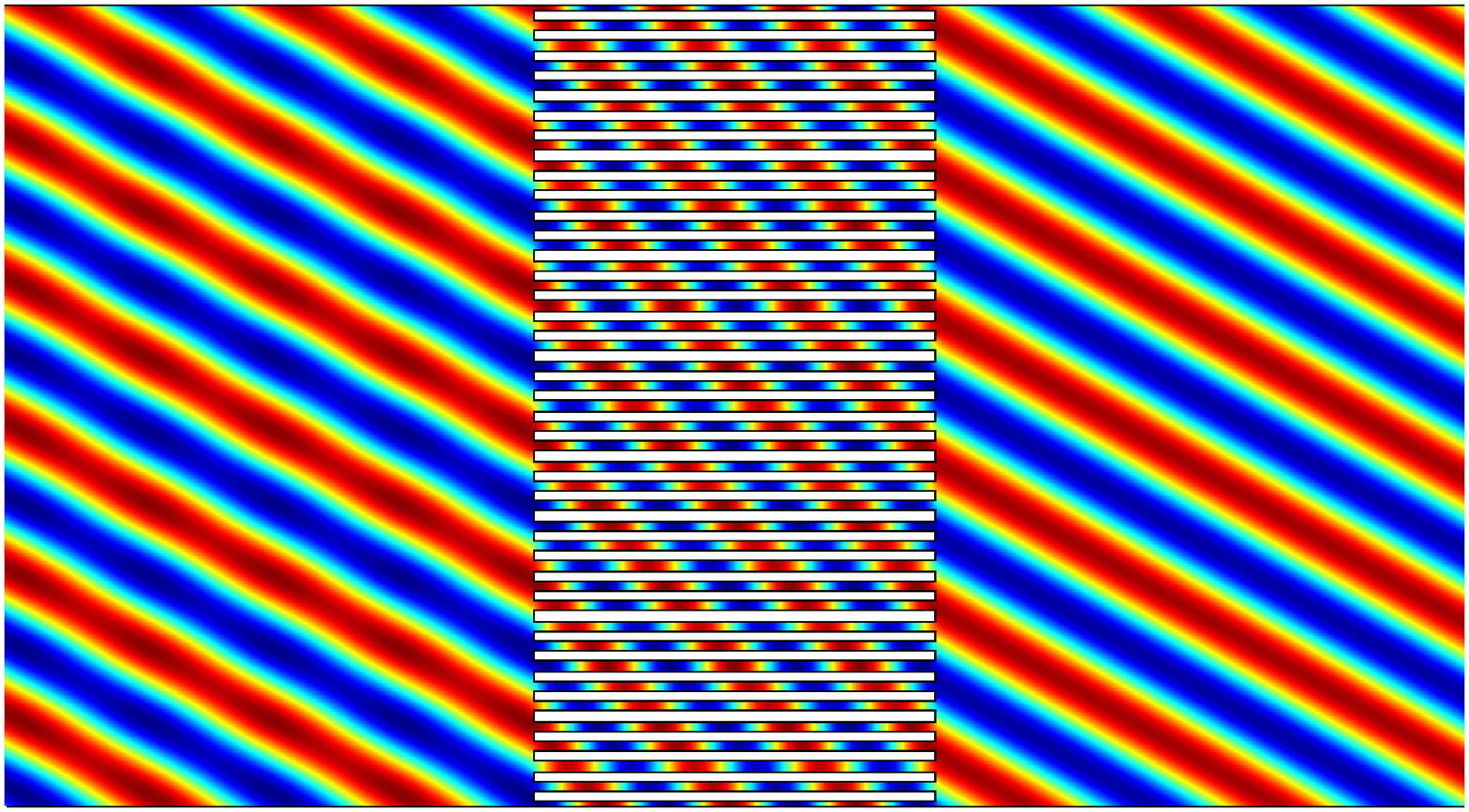} \label{pm0kd0752}}
\renewcommand{\thesubfigure}{(h)}\subfigure[$f=0.75$, $kd=1$]{
\includegraphics[width=0.34\textwidth]{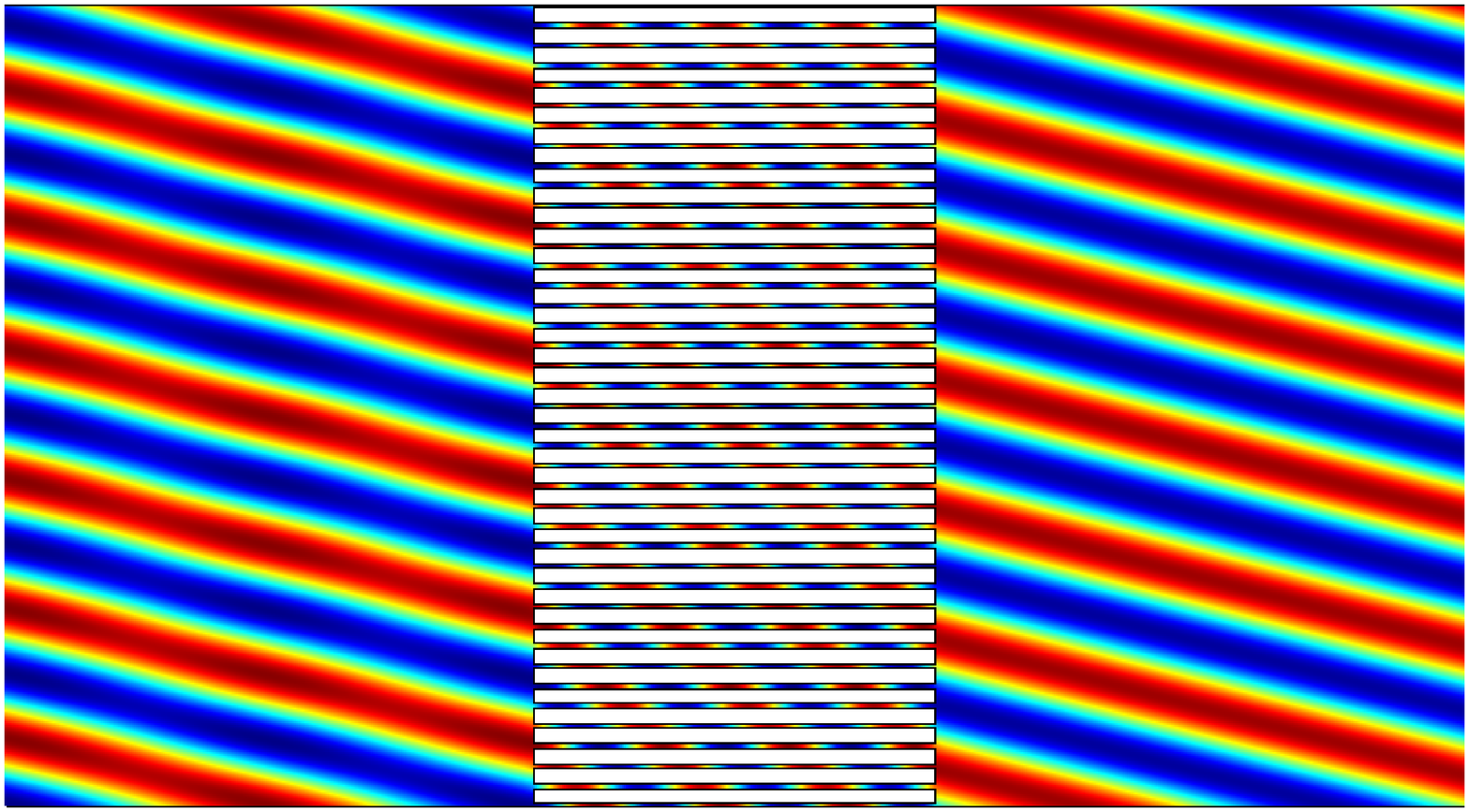} \label{pm0kd12}}
	\caption{Total pressure plots at different frequencies for a symmetric $(\phi = 0)$ slab of rigid elements.  The  incident angle is taken to be the intromission angle $\theta_i=\cos^{-1} (1-f)$, where $f=0.5$ $\Rightarrow \theta_i = 60^\circ$ in (a) through (d) while $f=0.75$ $\Rightarrow \theta_i = 41.4^\circ$ in (e) through (h).   The values of $kd$ range from $0.25$ to $1.0$, as indicated. 
	}   	\label{pm0}
\end{figure} 
\subsubsection{Rigid grating elements}
We begin with a symmetric slab of rigid elements, $\phi=0$,  in Figure \ref{pm0}.  The plots show  total pressure for  waves incident at the intromission angle for two different values of the filling  fraction, $f=0.5$ and $f=0.75$,  at  four frequencies at or below $kd=1$.  The plots  clearly show  total transmission for frequencies  $kd\ll 1$. 

 \begin{figure}[h] 
	\centering
\renewcommand{\thesubfigure}{(a)}\subfigure[$\phi=30^\circ$, $f=0.5$]{%
\includegraphics[width=0.34\textwidth]{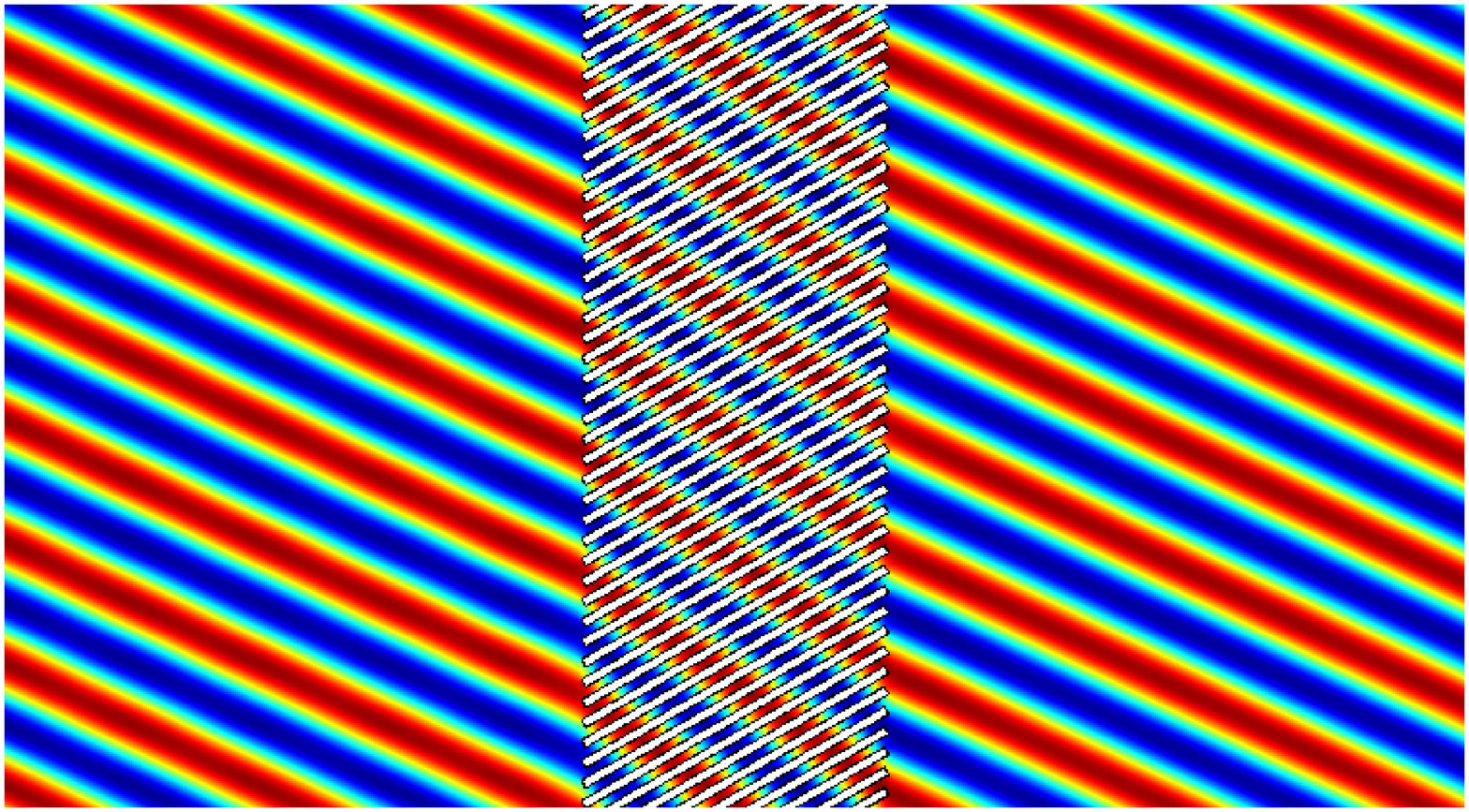}\label{pm30+}}
\renewcommand{\thesubfigure}{(c)}\subfigure[$\phi=30^\circ$, $f=0.75$]{%
\includegraphics[width=0.34\textwidth]{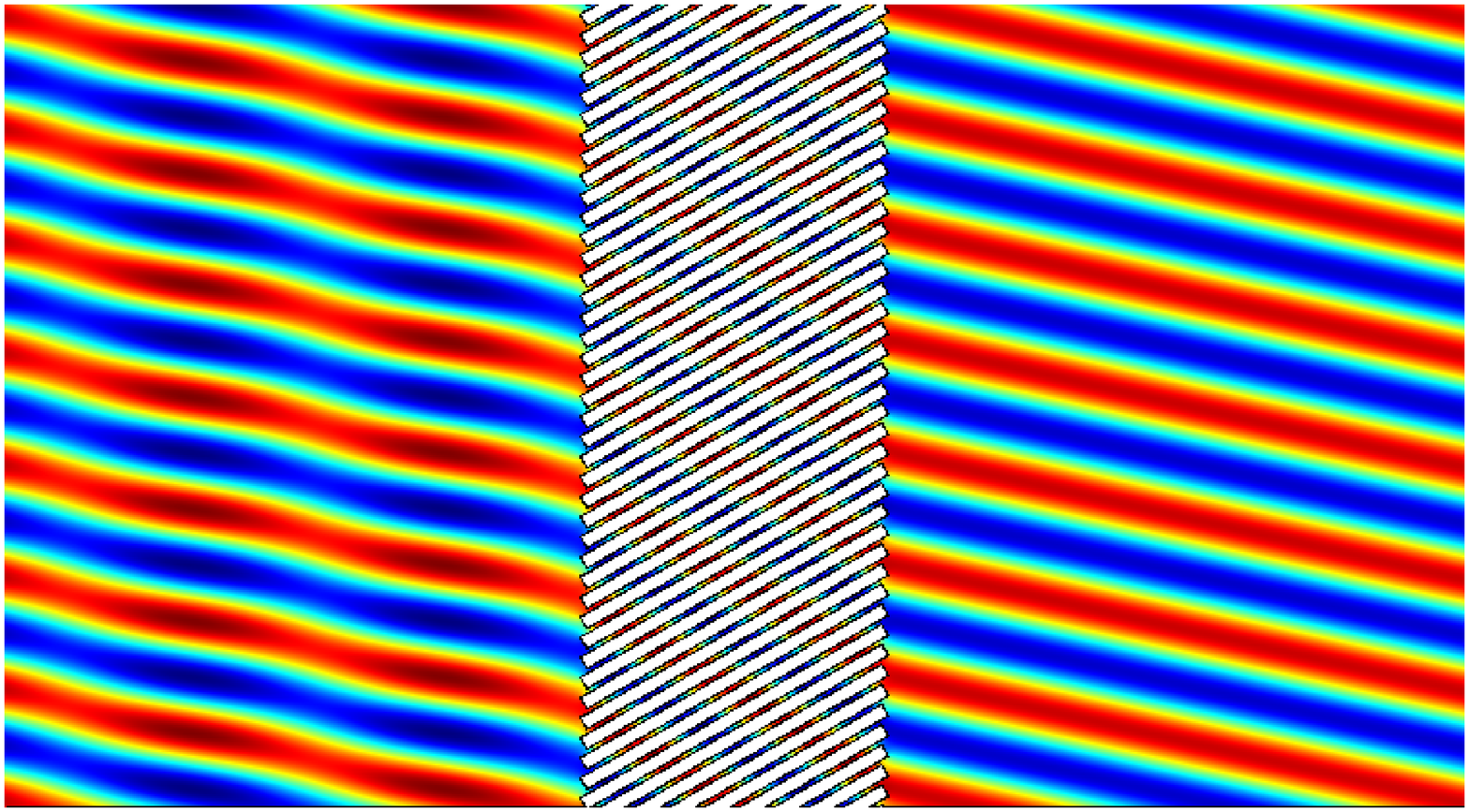}\label{pm30+2}}\\
\renewcommand{\thesubfigure}{(b)}\subfigure[$\phi=-30^\circ$, $f=0.5$]{%
\includegraphics[width=0.34\textwidth]{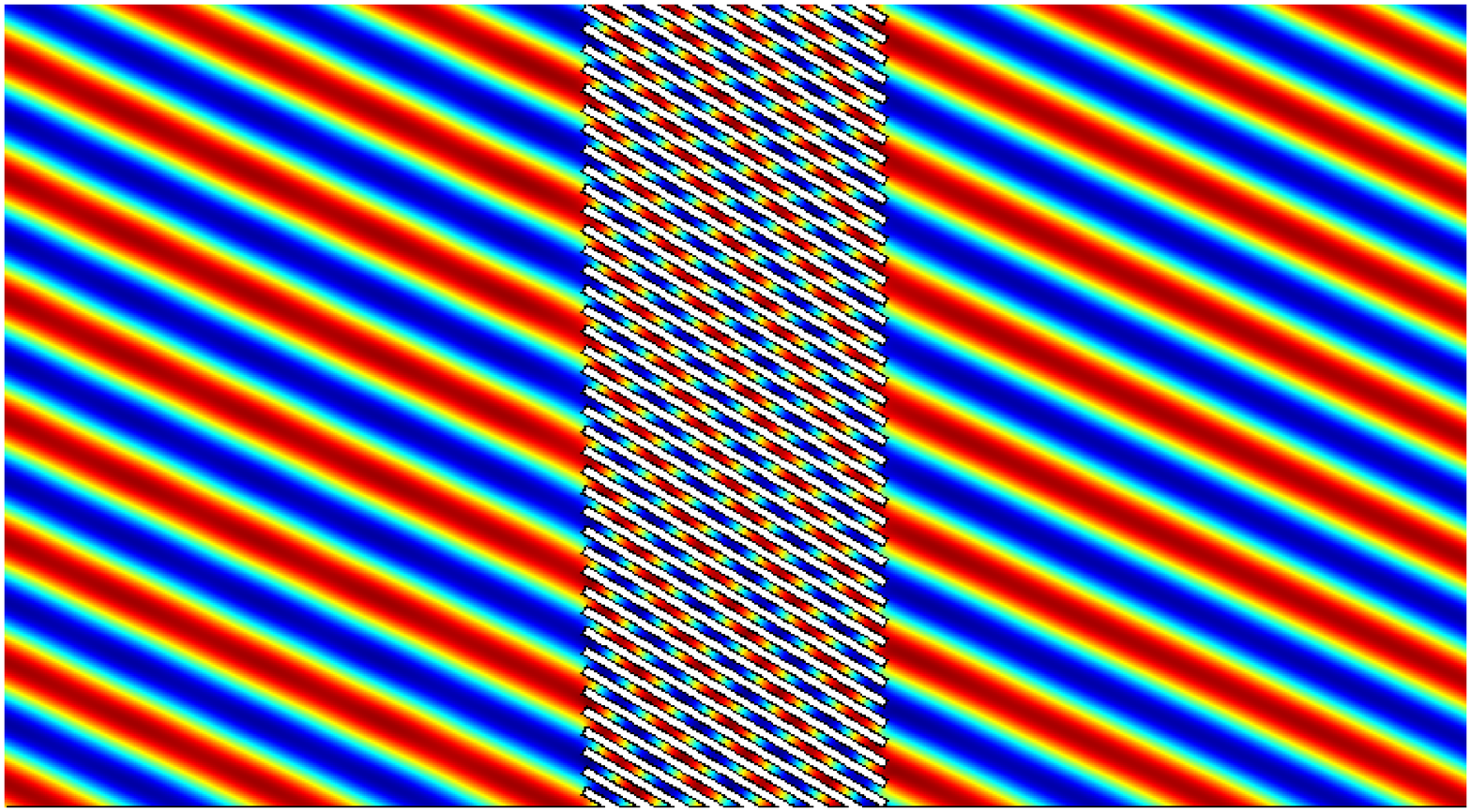}\label{pm30-}}
\renewcommand{\thesubfigure}{(d)}\subfigure[$\phi=-30^\circ$, $f=0.75$]{%
\includegraphics[width=0.34\textwidth]{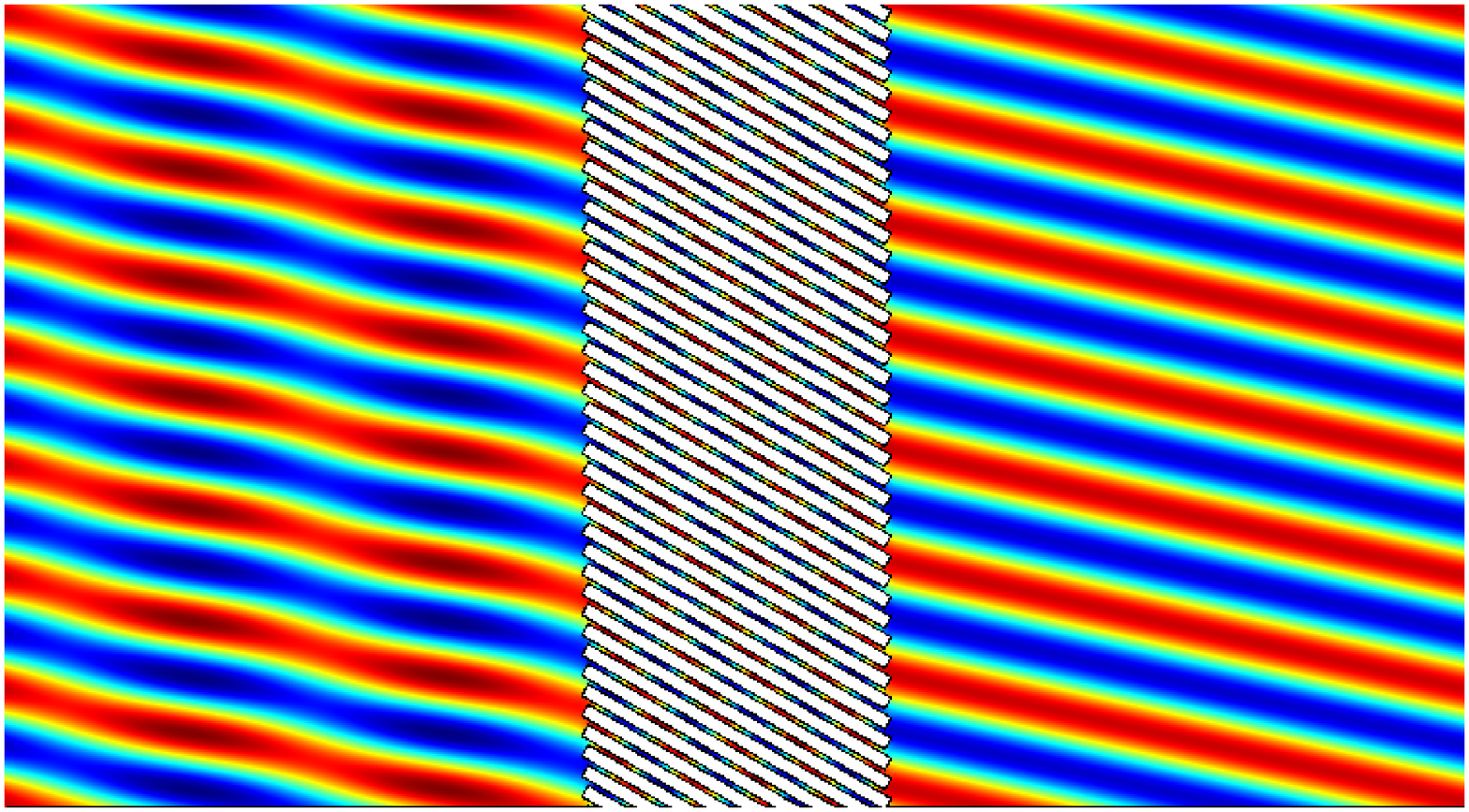}\label{pm30-2}}
	\caption{Total pressure plots for incidence at the intromission angle of a slanted grating of rigid elements, $\theta_i=\cos^{-1}\big ( (1-f)\cos \phi \big )$, $\phi=\pm 30^\circ$, at frequency $kd=1$.   
Plots (a) and (b) show the full pressure field for filling fraction $f=0.5$, while (c) and (d) are for higher filling fraction $f=0.75$.  
	}
	\label{pm30}
	\end{figure} 
\begin{figure}[H]
	\centering
	\renewcommand{\thesubfigure}{(a)}\subfigure[$\phi=60^\circ$, $kd=0.25$]{%
\includegraphics[width=0.34\textwidth]{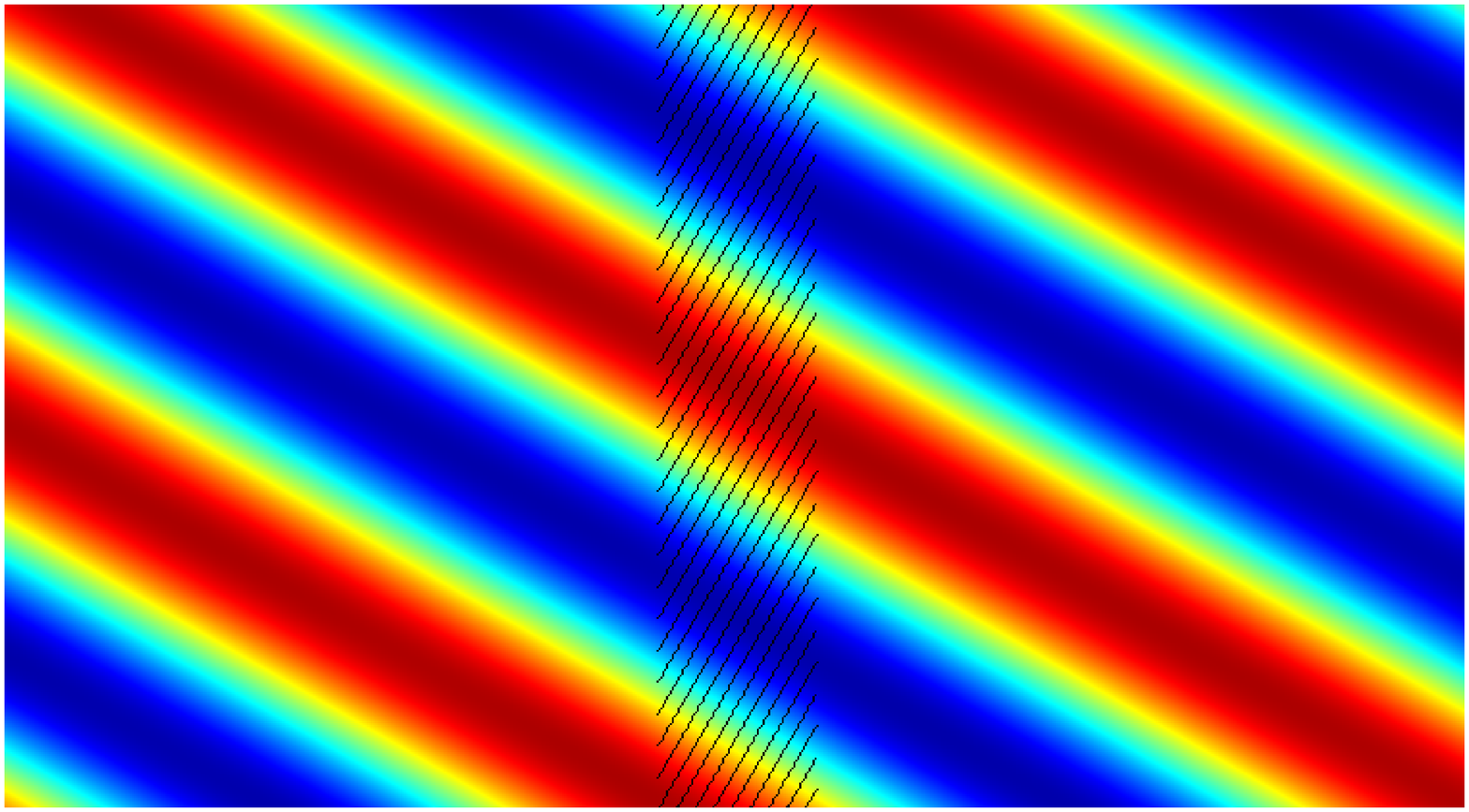}\label{pm01}}
\renewcommand{\thesubfigure}{(c)}\subfigure[$\phi=60^\circ$, $kd=0.5$]{%
\includegraphics[width=0.34\textwidth]{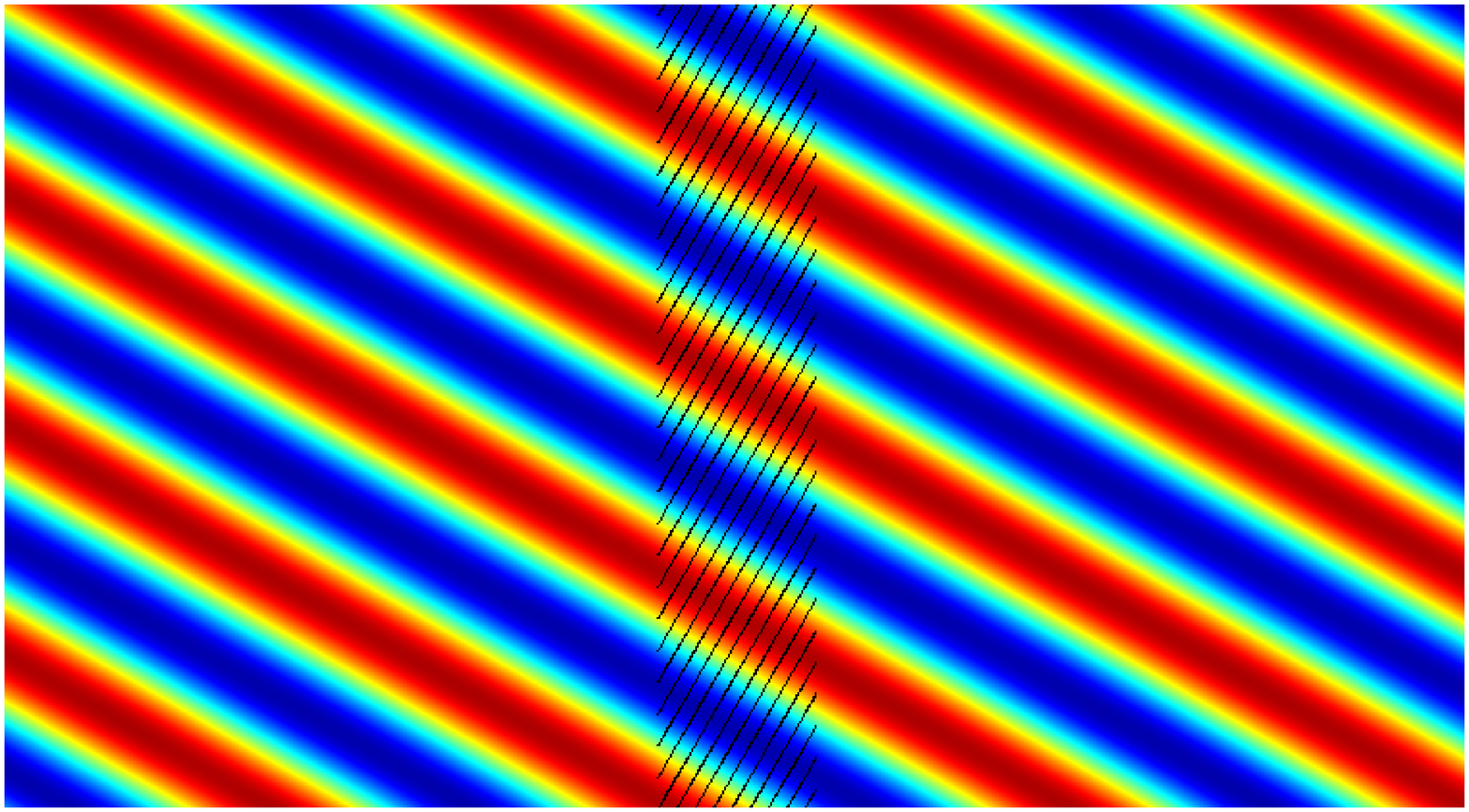}\label{pm03}}\\
\renewcommand{\thesubfigure}{(b)}\subfigure[$\phi=-60^\circ$, $kd=0.25$]{%
\includegraphics[width=0.34\textwidth]{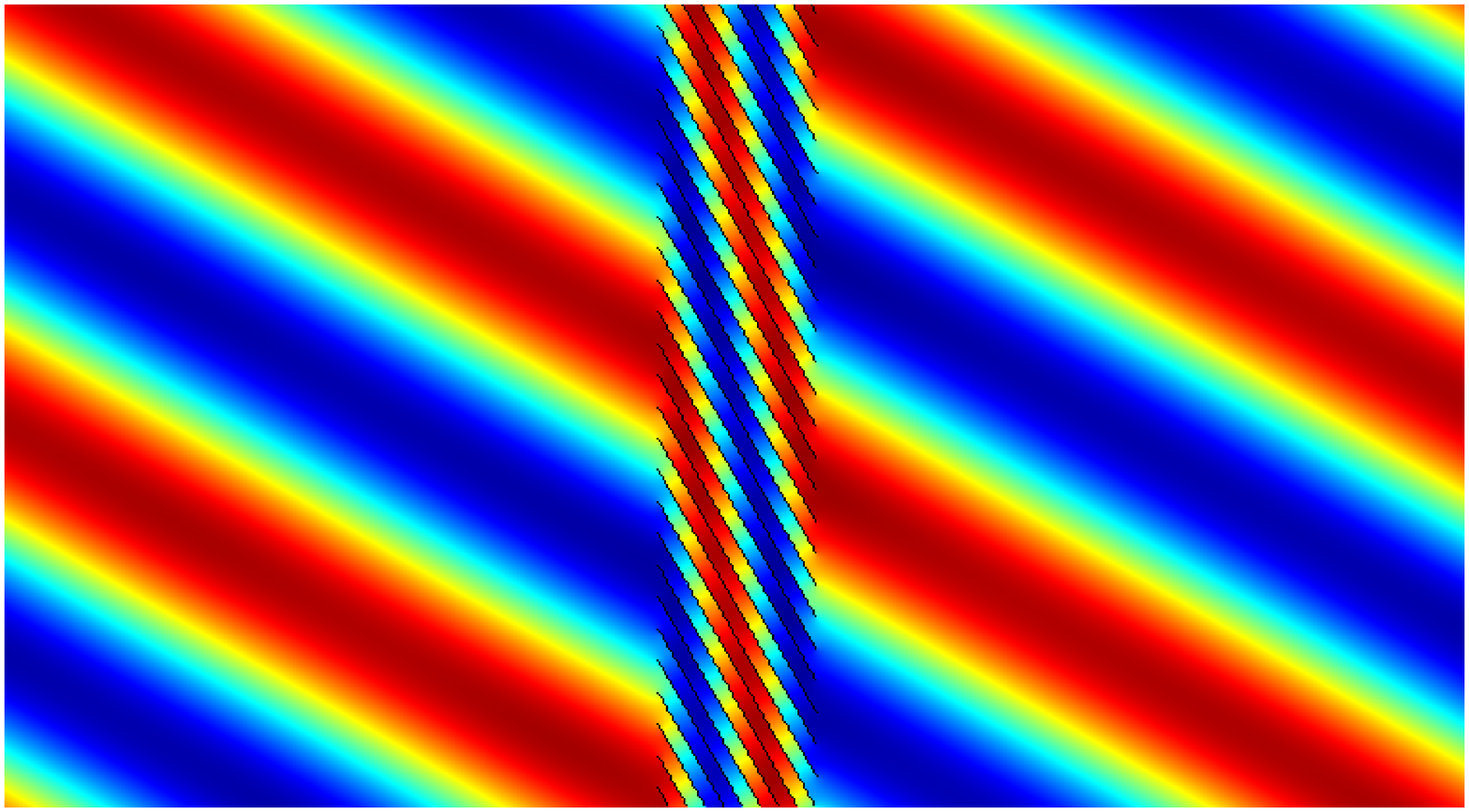}\label{pm02}}
\renewcommand{\thesubfigure}{(d)}\subfigure[$\phi=-60^\circ$, $kd=0.5$]{%
\includegraphics[width=0.34\textwidth]{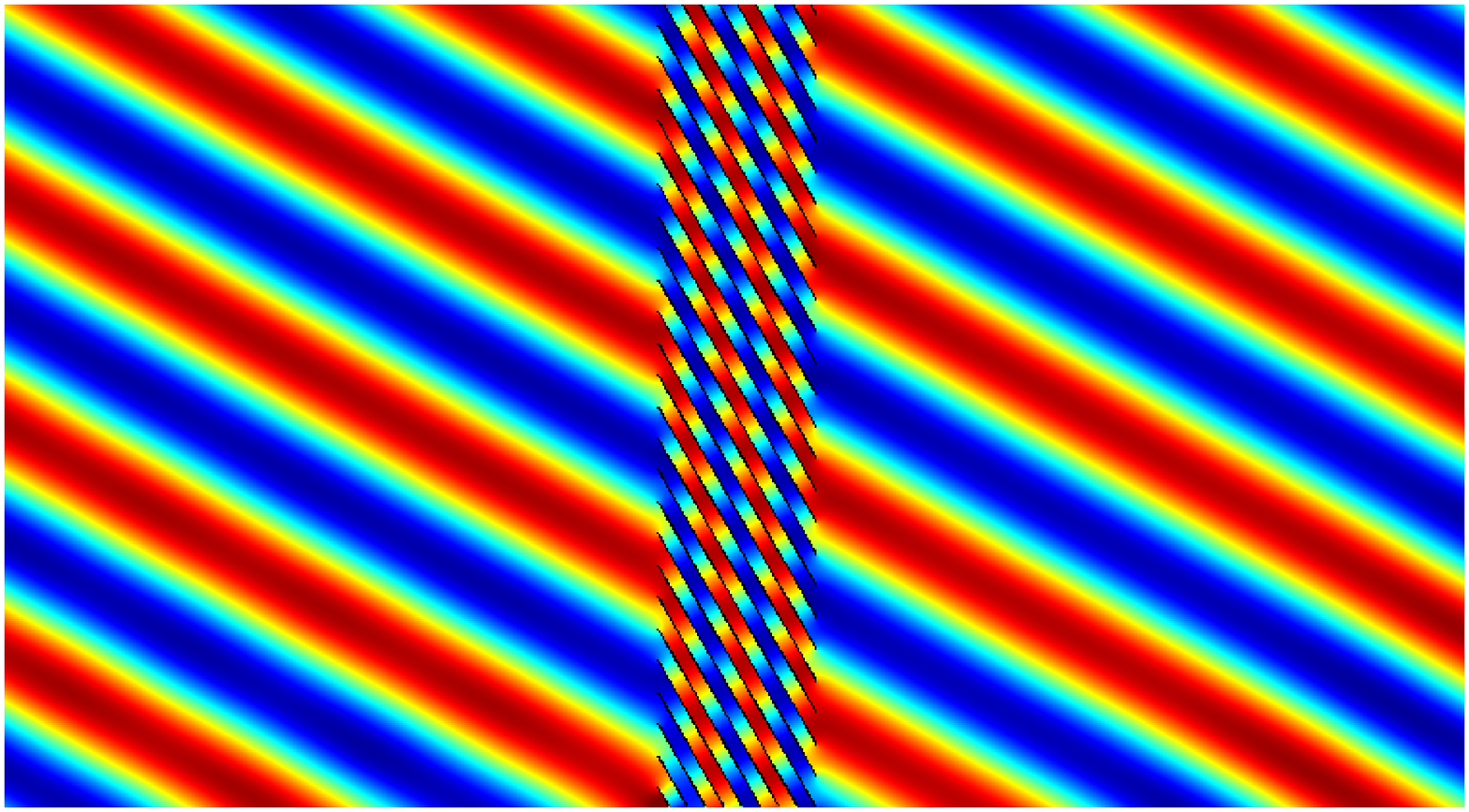}\label{pm04}}
	\caption{Wave transmission through a SLG of thin rigid elements (volume fraction  $f=0+$) 
	oriented at  $\phi=\pm60^\circ$ for incidence at the intromission angle $\theta_i=60^\circ$.}
	\label{0pm0}
\end{figure}
\begin{figure}[h] 
\centering
\includegraphics[width=0.6\textwidth]{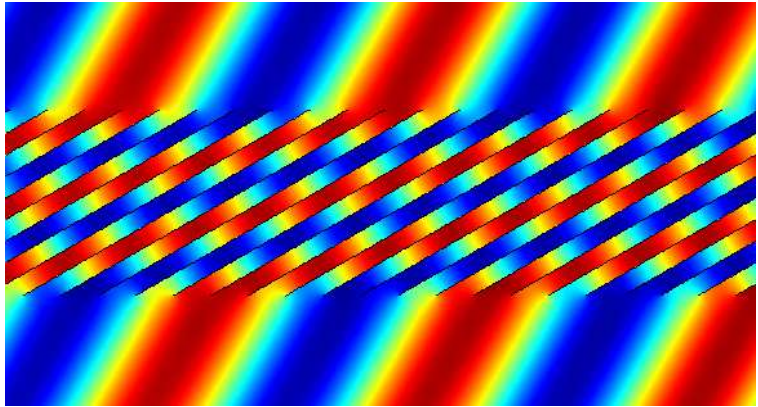}\label{rigid-}
	\caption{Zoomed-in view of the phase difference in plot  (b) of  Figure \ref{0pm0}. }
	\label{rigid}
\end{figure}

 
As noted above, full transmission through an asymmetric grating of elements oriented at angle $\phi$ can be obtained at both $\theta_i$ and $-\theta_i$. In the numerical experiments shown in Fig.\   \ref{pm30} we change the direction of  rotation of  slab elements instead of changing the incident direction, i.e., using $\pm\phi$ instead of $\pm\theta$. Figure \ref{pm30} shows that the pressure amplitude  transmitted through the slab for $\phi$ is the same as  for $-\phi$. The transmitted phases  are clearly different; the phase effect is easier to see in Fig.\   \ref{pm30} for the example with lower filling fraction.   In  the limit of zero but still finite filling fraction,
$f=0+$,  the rigid element SLG acts like a comb, totally transparent for incidence at $\theta_i = \pm \phi$, as illustrated in  Figure \ref{0pm0} with a zoom-in shown in Fig.\  \ref{rigid}.  The phase difference between incidence at 
$\theta_i = + \phi$ and $\- \phi$ is most dramatic in this limit of thin rigid grating elements.   This proves the original assertion about what happens in Fig.\  \ref{line}(b).


\begin{figure}[h] 
	\centering
\includegraphics[width=5.in, height=3.in]{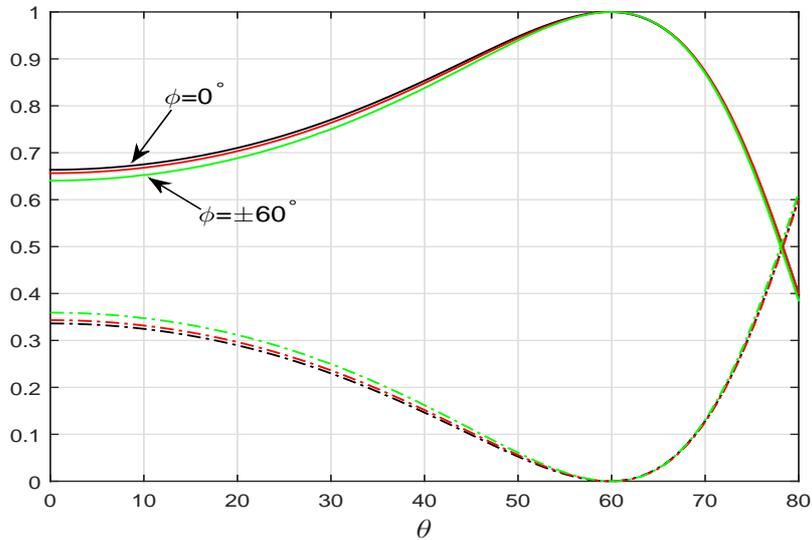}
	\caption{Full transmission at $\theta_i=60^\circ$ for three different  rigid SLGs with elements oriented at $\phi=0^\circ$, $30^\circ$ and $60^\circ$.  The solid and dashed curves show $|T|^2$ and $|R|^2$, respectively. The black, red, and green curves are for the cases $\phi=0^\circ$, $\pm30^\circ$ and $\pm60^\circ$, respectively.  The frequency is $kd=0.25$. } 
	\label{trans}
\end{figure}

As a final example of a grating with rigid elements, Fig. \ref{trans} 
shows the computed reflection and transmission coefficients for three different slanted gratings.  The intromission angle in each case was chosen  to be $\theta_i=60^\circ$ which constrains 
the orientation angle $\phi$ and the volume fraction $f$ to satisfy $(1-f)\cos\phi = \frac 12$, see eq.\  \eqref{67-2}.   Figure \ref{trans} indicates that the transmission spectrum does not change significantly  as long as the relation between  $\theta_i$, $\phi$ and $f$ is obeyed.   

\subsubsection{Acoustic grating elements}

\begin{figure}[h] 
	\centering
\includegraphics[width=0.75\textwidth]{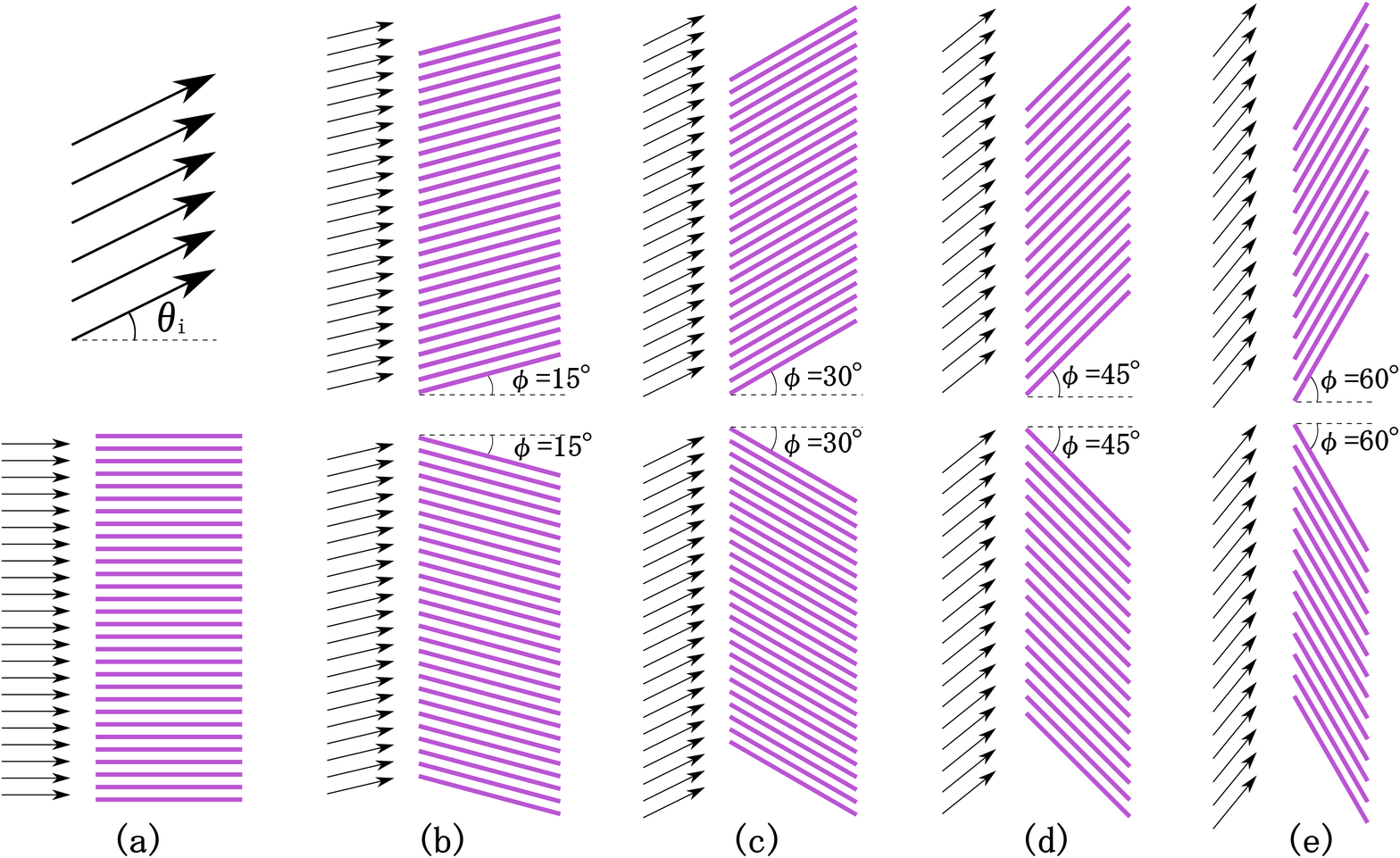}
	\caption{The intromission angle changes with $\pm\phi$ for an acoustic grating with filling fraction $f=0.3$ subject to the constraint that  (a) $\theta_i=0^\circ$ for $\phi = 0$.  The intromission angle $\theta_i$ in the other cases is: 
	(b) $13.3^\circ$, 
	(c) $26.3^\circ$, 
	(d) $38.8^\circ$,
	(e) $50.1^\circ$.} 
	\label{rotate}
\end{figure}

\begin{figure}[h] 
\centering
\renewcommand{\thesubfigure}{(a)}\subfigure[$\phi=30^\circ$]{%
\includegraphics[width=0.34\textwidth]{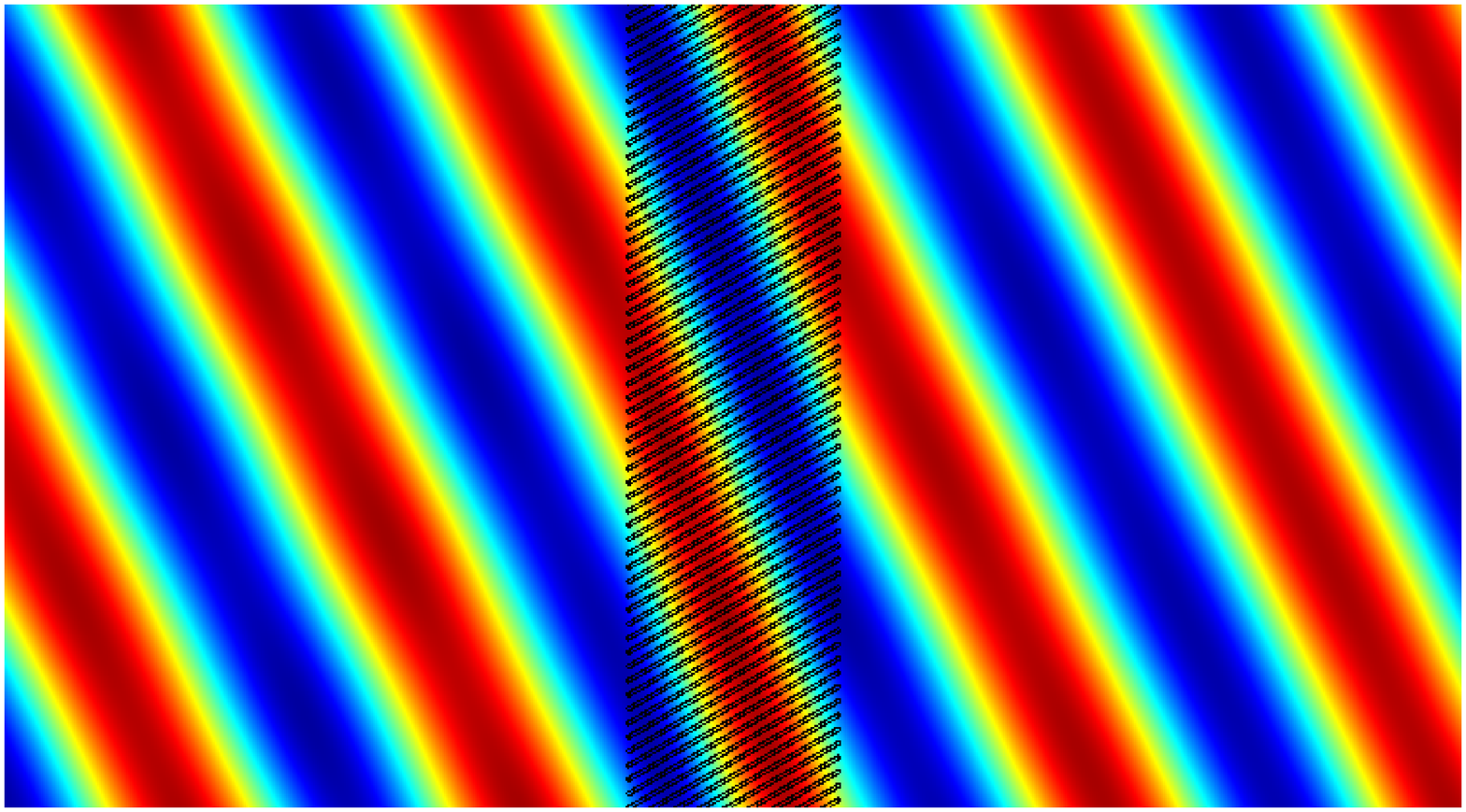}\label{nonrigid1+}}
\renewcommand{\thesubfigure}{(c)}\subfigure[$\phi=60^\circ$]{%
\includegraphics[width=0.34\textwidth]{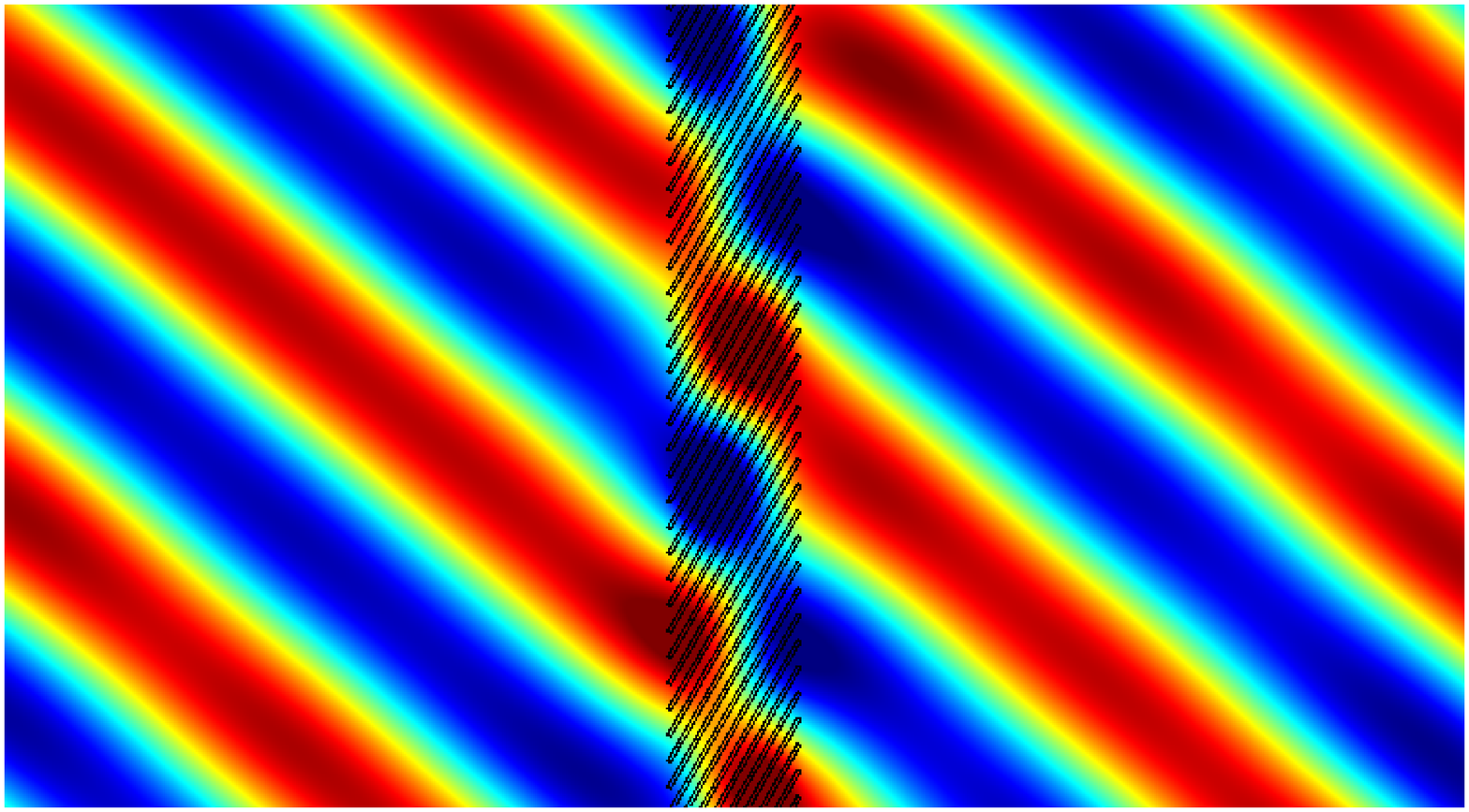}\label{nonrigid2+}}\\
\renewcommand{\thesubfigure}{(b)}\subfigure[$\phi=-30^\circ$]{%
\includegraphics[width=0.34\textwidth]{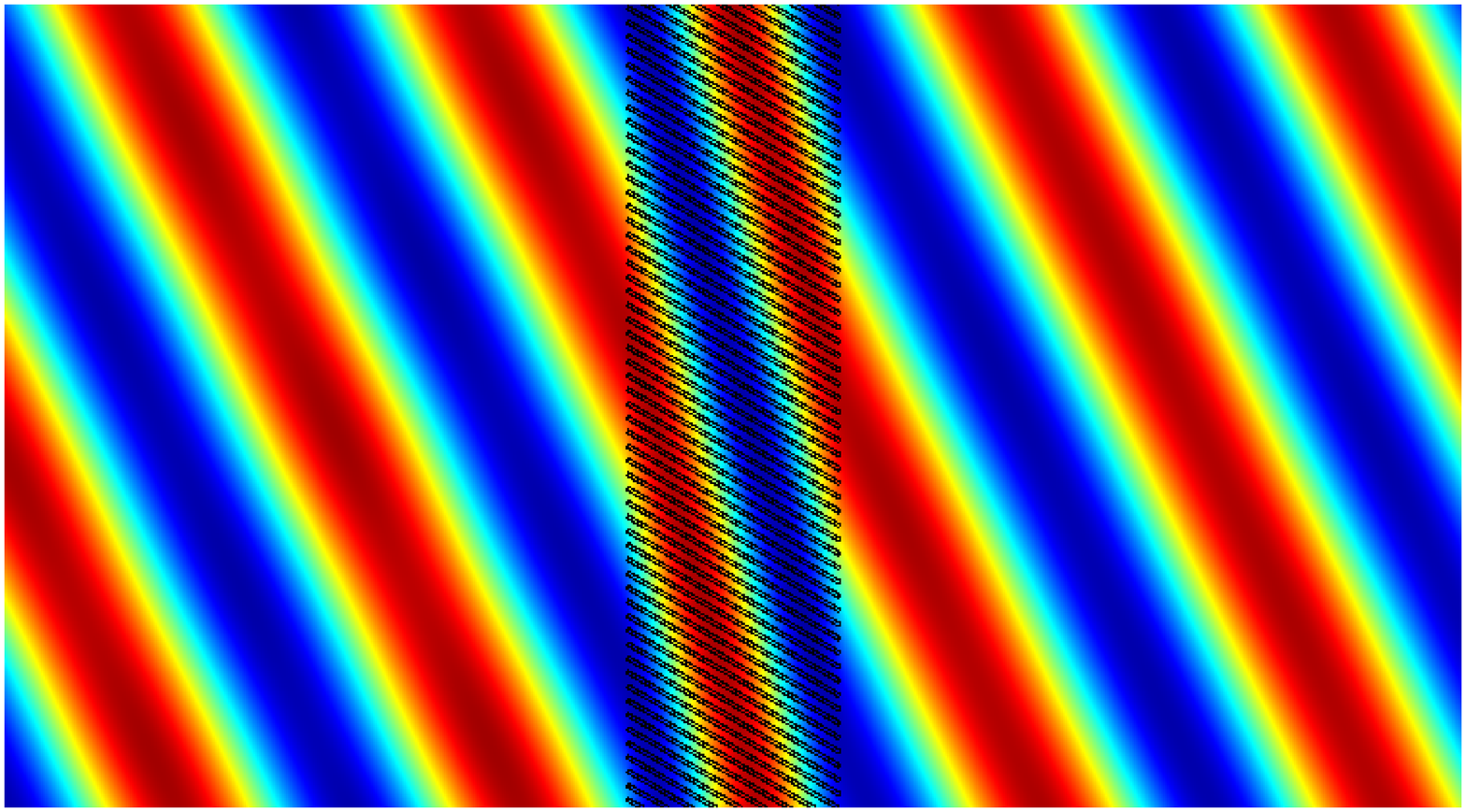}\label{nonrigid1-}}
\renewcommand{\thesubfigure}{(d)}\subfigure[$\phi=-60^\circ$]{%
\includegraphics[width=0.34\textwidth]{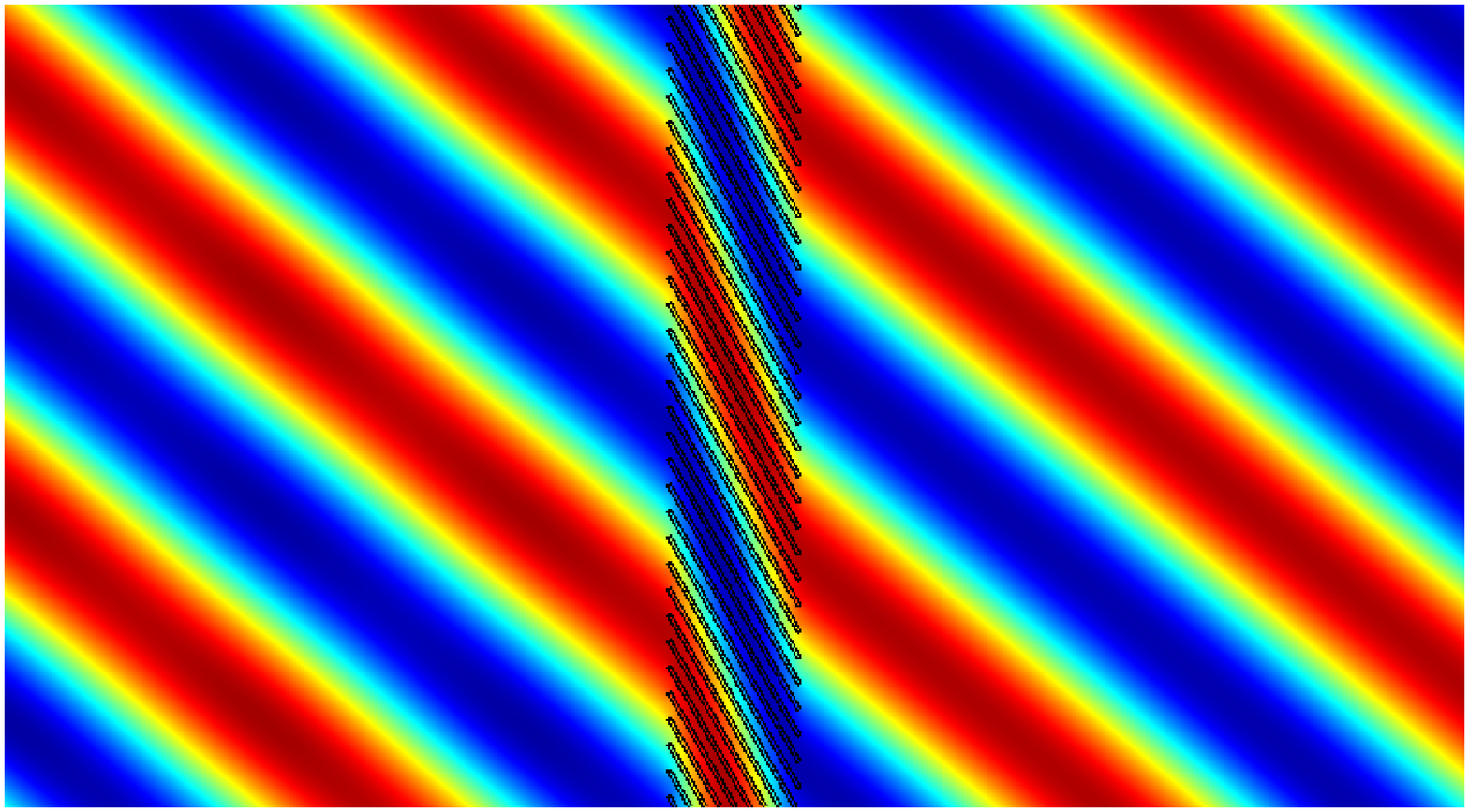}\label{nonrigid2-}}
	\caption{Total pressure plots for some  configurations from  Fig.\  \ref{rotate} at frequency $kd=0.25$. }
	\label{nonrigid}
\end{figure}

The material properties of the gratings are 
selected so that the intromission angle is zero when the grating elements are symmetric, i.e.\ $\theta_0=0$ for $\phi = 0$, and we consider the change in properties as the elements are subsequently rotated to $\phi > 0$.  
The  background  acoustic medium is assumed to be water, $\rho = 1000$ kg/m$^3$, $c=1500$ m/s, and $K=2.25$ GPa. 
We first consider  a denser fluid, 
 $\rho_0=10\rho$, at volume fraction $f=0.3$, then    equation \eqref{74} yields $K_0=1.008$  GPa, guaranteeing that a wave of normal incidence has full transmission for $\phi=0^\circ$.  We  then vary $\phi$, with all other material parameters fixed, to calculate the intromission angle 
for each $\phi$ according to eq.\ \eqref{75}. Figure \ref{rotate} shows how the intromission angle changes with $\phi$. Notice that $\phi$ can be positive or negative so that the gratings can rotate in two directions. The full field  shown in Fig.\  \ref{nonrigid} illustrates  the phase transfer across the SLG.  This is clearly   more complicated than in the rigid case, where the acoustic propagation is along parallel waveguides.     The interaction of the two fluids in the SLG is particularly evident in
Fig.\ \ref{nonrigid}(c).  

\begin{figure}[h] 
	\centering
\includegraphics[width=0.75\textwidth]{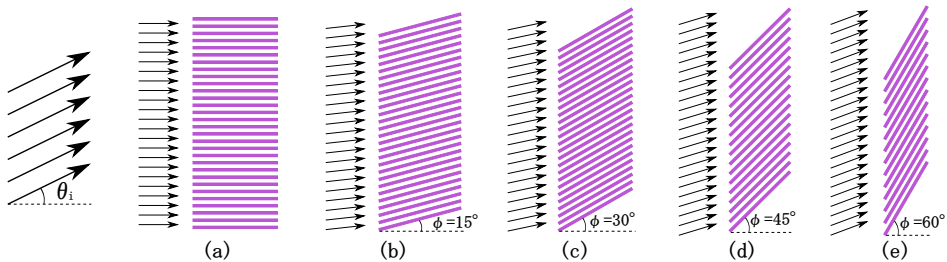}
	\caption{The intromission angle  is constrained to be zero for symmetric alignment $(\phi = 0)$ for a grating with fluid less dense and stiffer than the background, (a).   For filling fraction $f=0.3$, the grating elements are rotated by angle  $\phi$ and  the intromission angle $\theta_i$ becomes 
	(b) $6.5^\circ$,
	(c) $12.5^\circ$,
	(d) $17.6^\circ$,
	(e) $21.4^\circ$.} 
	\label{rotate2}
\end{figure}

\begin{figure}[h] 
\centering
\renewcommand{\thesubfigure}{(a)}\subfigure[$\phi=30^\circ$]{%
\includegraphics[width=0.34\textwidth]{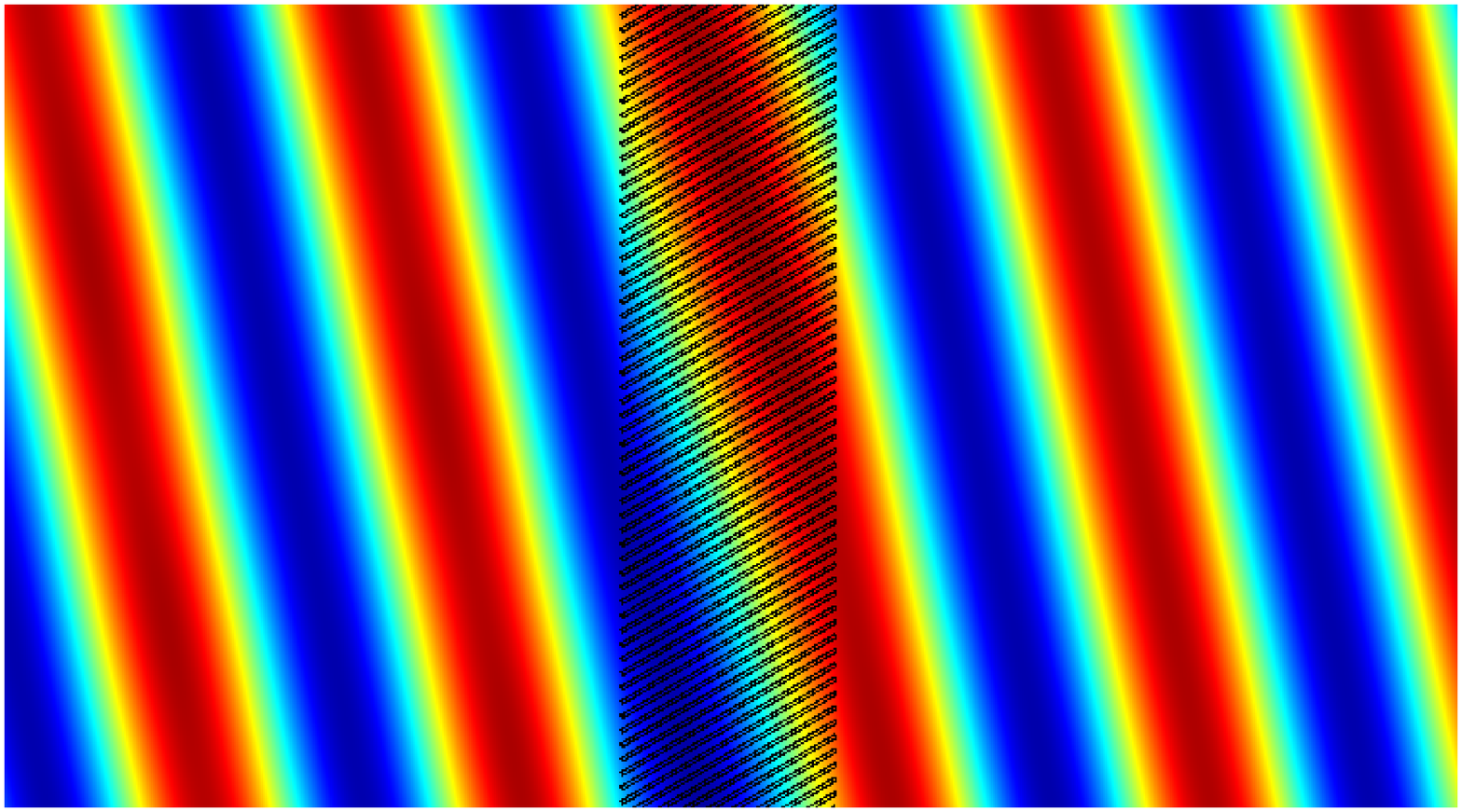}\label{nonrigid3+}}
\renewcommand{\thesubfigure}{(c)}\subfigure[$\phi=60^\circ$]{%
\includegraphics[width=0.34\textwidth]{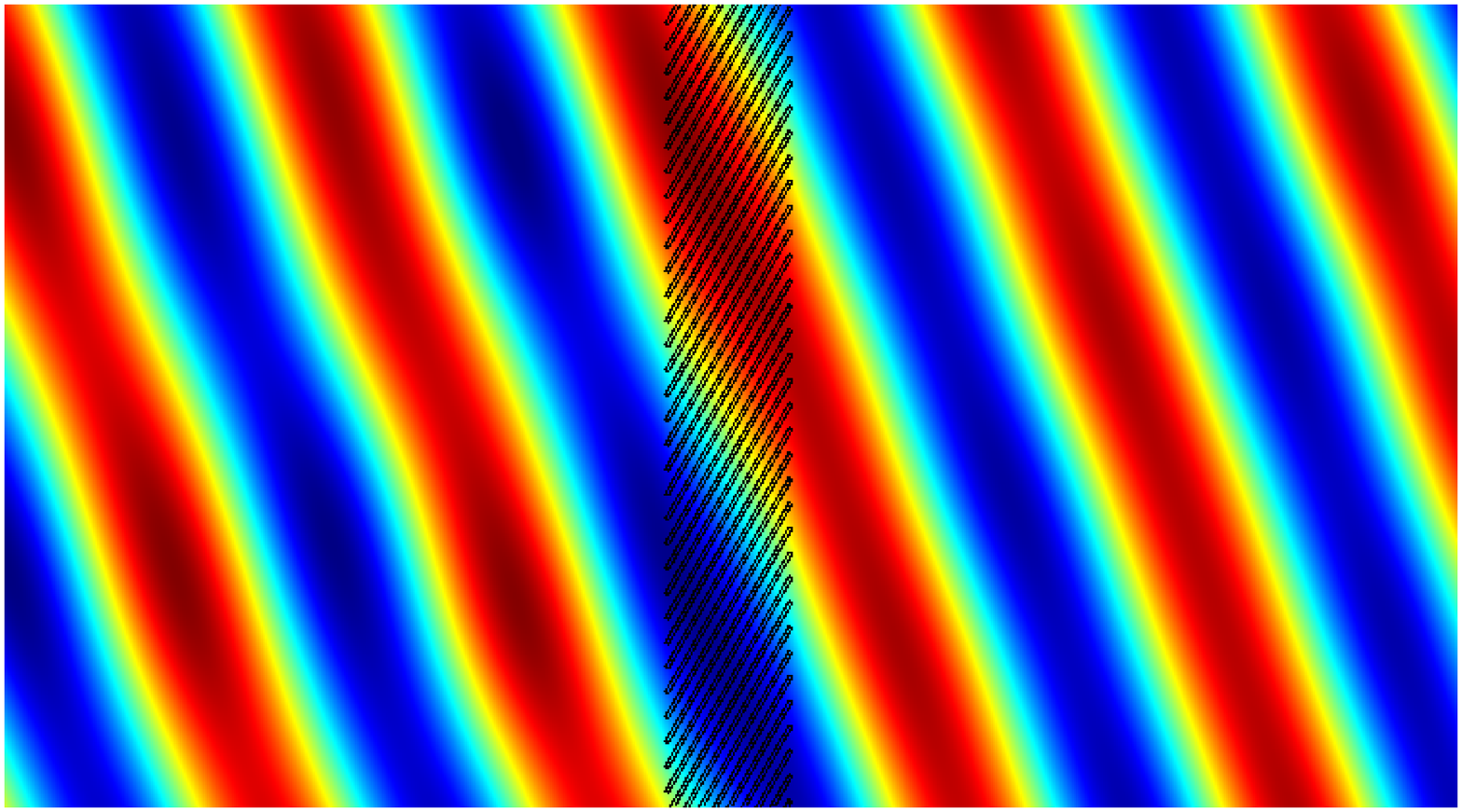}\label{nonrigid4+}}\\
\renewcommand{\thesubfigure}{(b)}\subfigure[$\phi=-30^\circ$]{%
\includegraphics[width=0.34\textwidth]{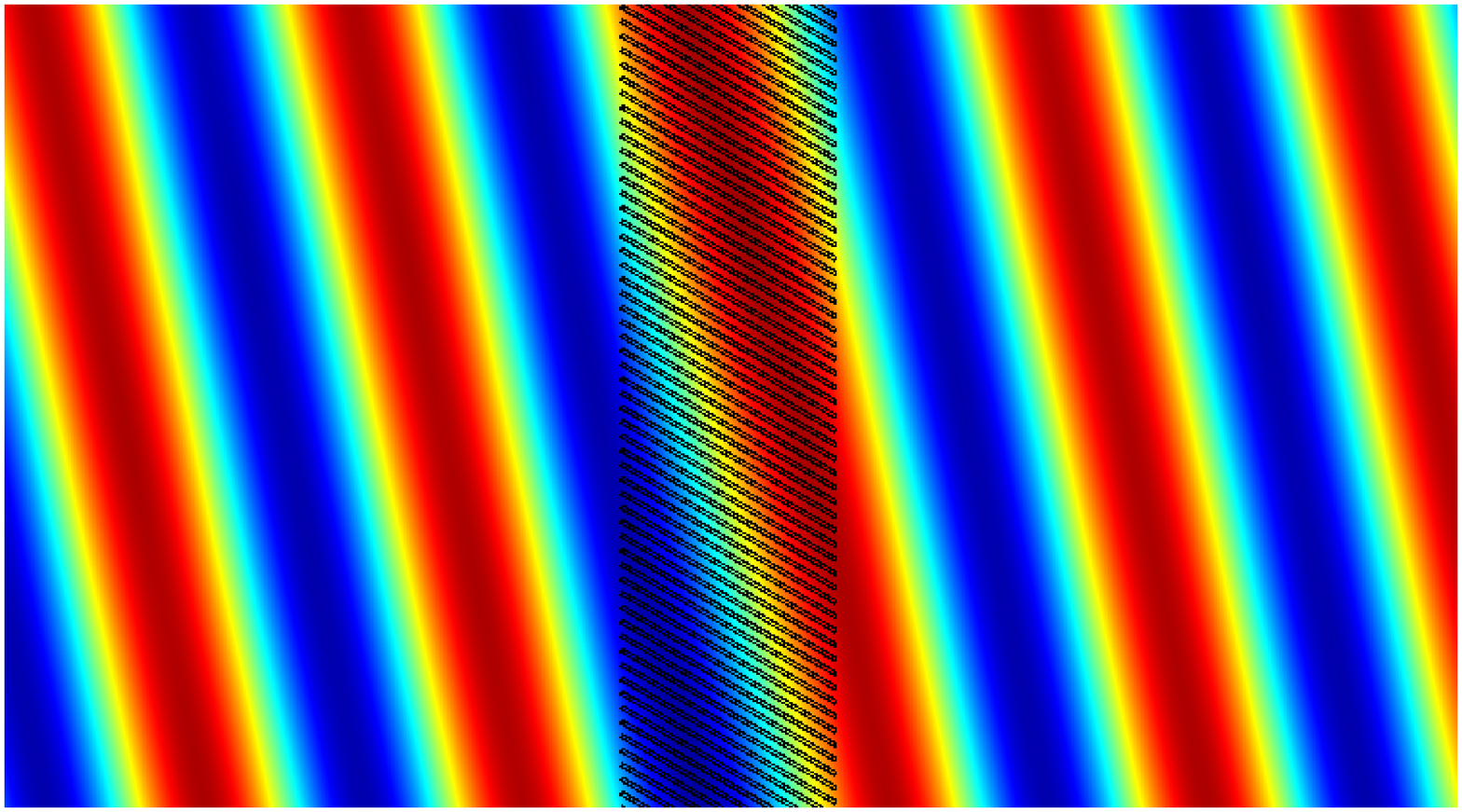}\label{nonrigid3-}}
\renewcommand{\thesubfigure}{(d)}\subfigure[$\phi=-60^\circ$]{%
\includegraphics[width=0.34\textwidth]{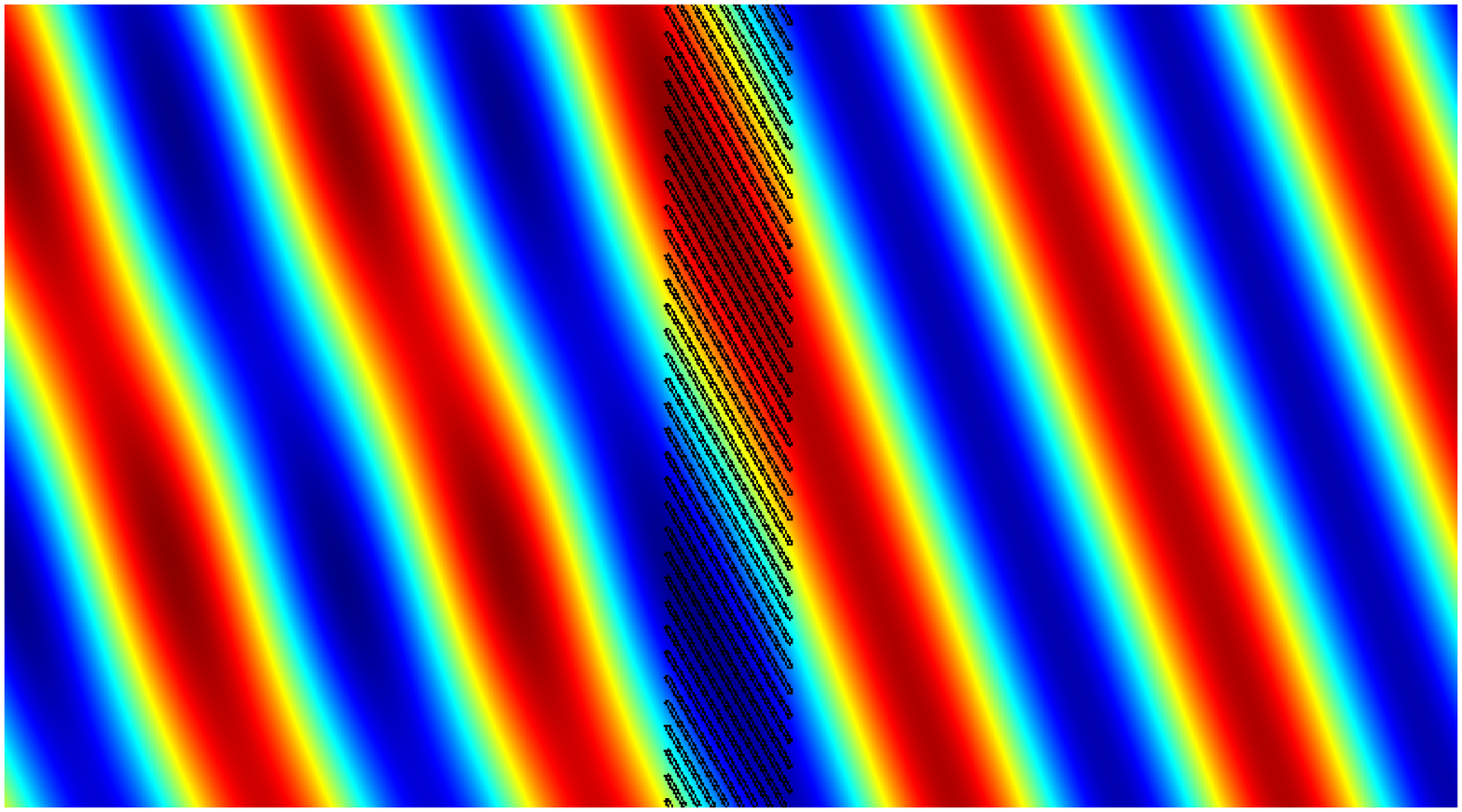}\label{nonrigid4-}}
	\caption{Total pressure plots for some  configurations from  Fig.\  \ref{rotate2} at frequency $kd=0.25$. }
	\label{nonrigid_new}
\end{figure}

In the   previous example   the grating elements   were chosen as denser than the background (water) and it was found that the grating had to be less stiff (lower bulk modulus) than water. Conversely, if we consider 
a SLG using a fluid that is lighter than the background, the same constraint that 
the intromission angle is zero for symmetric alignment, $\theta_i=0$ at $\phi = 0$,
requires that the fluid is  stiffer than water.  For instance, eq.\  \eqref{74} is satisfied with $f=0.3$,  $\rho_0=0.45\rho$ and $K_0=21.29$  GPa, so that the intromission angle is zero for the symmetric   configuration $\phi = 0$.  Figure \ref{rotate2} shows how the intromission angle changes with $\phi$ for this grating. 
Figure \ref{nonrigid_new} show the full field for  $\phi=\pm 30^\circ$ and  $\phi=\pm 60^\circ$.    It is instructive to compare these results with those for the other fluid in 
Fig.\ \ref{nonrigid}. 

It is possible, in principle, to design materials with low density $\rho_0 < \rho$ and high stiffness $K_0>K$.  Metal foams, e.g. Duocel\textregistered aluminum foam, can have very low density $\rho_0 \ll  \rho$ and relatively high stiffness, however, the random structure usually limits the effective bulk modulus to be less than that of water.  Simultaneously ultra-light and ultra-stiff materials are obtained using thin lattice structures with ordered periodicity \cite{Fang14}.  These materials possess significant shear modulus, i.e.\ the Poisson's ratio is not close to $\frac 12$, which implies they support both shear and longitudinal waves.   By carefully selecting the unit cell one can achieve a one-wave fluid like material with properties $\rho_0=\rho$ and $K_0=K$ of water, specifically known as Metal Water \cite{Norris11mw}.   The low shear rigidity is ensured by using very thin members with large flexural compliance.    The metal water structure, designed to have  quasistatic properties of water, also exhibits interesting band structure which makes it a narrow-band negative index  material \cite{Hladky-Hennion13}.  The generalization of the metal water structure is a class of metallic pentamode materials with  low-shear and design specific density and stiffness, which  could in principle achieve desired values of $\rho_0$, $K_0$.

\section{Conclusions}\label{sec6}

Our main result is eq.\ \eqref{67} which gives the intromission angle for the single-layer grating of Fig.\ \ref{fig2}.  While it is known that EAT can be understood as impedance matching  in the context of acoustics of fluids with anisotropic inertia \cite{Maurel13} the present results show that this analogy extends further to include  asymmetric gratings.   The  principal axes of the  anisotropic inertia are  not necessarily aligned with the slab axes (Fig.\ \ref{fig4}) which introduces asymmetry in the phase of the transmitted wave as a function of incidence angle $\pm \theta$.    These seemingly unusual results for total transmission 
can be easily understood when the grating elements are rigid.  Thus, any angle of intromission can be obtained with thin rigid elements by orienting them to the desired value of $\theta_i$, a simple comb-like effect.   Surprisingly, full transmission is also achieved at incident angle $-\theta_i$, see Fig.\ \ref{0pm0}.  

The  rigid grating with thin slanted elements  illustrates the geometrical acoustics nature of the EAT phenomenon. However, the simultaneous EAT effect at orientations $\pm\phi$
emphasizes that the underlying phenomenon is ``geometrical impedance matching''.  The term geometrical impedance matching is introduced to signify the flux condition across the interface, as compared with the phase matching (Snell's or Descartes' law) in the transverse direction.  Thus,  geometrical impedance matching leads directly to the identity \eqref{67-2} for the rigid grating.   However,  one needs a full wave approach in order to arrive at the more general result of eq.\ \eqref{67} 
for the intromission angle in the presence of an acoustic fluid grating. Despite this, the   simplicity  of the identity  eq.\ \eqref{67}  for the intromission angle is remarkable.

\appendix 

\Appendix{Solution for an anisotropic inertial slab} \label{321}

The slab properties are bulk modulus $K_s$ and 2$\times$2 inertia matrix 
$\boldsymbol{\rho} =\boldsymbol{\rho}^T$. 
Define the state vector 
\beq{201}
{\bf u}= \begin{pmatrix} v_1  \\ -p \end{pmatrix}, 
\eeq
and consider solutions with  constant horizontal phase such that  ${\bf u}$ has the form 
\beq{14}
{\bf u}(x_1,x_2)={\bf U}(x_1)e^{ik \sin \theta \, x_2  } . 
\eeq
Then    ${\bf U}(x_1)$ satisfies 
\bal{-021}
\frac{\dd {\bf U} }{\dd x_1} &= i\omega {\bf A} {\bf U} 
\ \ \text{where} 
\\
{\bf A}  &= \frac{\sin \theta}c \frac{\rho_{12}}{\rho_{22}} {\bf I}-{\bf B}, 
\quad
{\bf B}  = 
\begin{pmatrix}
 0 & \frac 1{K_s} - \frac{\sin^2 \theta}{c^2 \rho_{22}}  
\\ 
 \frac{\det \boldsymbol{\rho}}{\rho_{22}}  
&  0
\end{pmatrix}
\label{-45} 
\eal
 and ${\bf I}$ is the identity matrix. Note that the matrix ${\bf A}$ is   independent of the frequency $\omega$. 

Define the propagator matrix,  ${\bf M}   (x) $,   
as the solution of
\beq{-323}
\frac{\dd {\bf M}  (x) }{\dd x} = { i\omega} {\bf A} {\bf M} 
\ \ \text{with} \ {\bf M}(0) = {\bf I}.
\eeq
Note that $\det {\bf M} =1$  \cite{Norris10}.  The property    ${\bf A}^T = {\bf J} {\bf A} {\bf J} $ where the 2$\times$2 matrix ${\bf J}$ has zeroes on the diagonal and   unity  off diagonal,  implies that the Hermitian conjugate satisfies 
${\bf M}^\dagger = {\bf J} {\bf M}^{-1} {\bf J} $ and hence 
${\bf M}^{-1} (x)  = {\bf J} {\bf M}^\dagger (x) {\bf J} 
= {\bf M}(-x)$. 
 
We  consider slabs with   uniform properties in $x\in [0,\a]$, so that 
\beq{313}
{\bf M}  (\a) = e^{ { i\omega } \a {\bf A} }. 
\eeq
This explicit form of the propagator matrix  simplifies, using eqs.\ \eqref{-45} and  \eqref{313}
and the property that $c_\theta^{-1}{\bf B}$ is a square root of the identity, to  give
\beq{61}
{\bf M}(\a) = 
\Big(
\cos \frac{\omega \a}{c_\theta}    \,{\bf I}  - \frac{i}{c_\theta} \sin\frac{\omega \a}{c_\theta}  \,{\bf B}  
\Big)\, e^{ik \a \frac{\rho_{12}}{\rho_{22}}\sin\theta },
\eeq
where
\beq{62}
c_\theta = \big(-\det {\bf B}\big)^{-1/2}. 
\eeq

Based on eqs.\ \eqref{-1} and \eqref{14}, 
\beq{91}
{\bf U}(0-)  = p_0 \begin{pmatrix}
Z_\theta^{-1} (1-R)
\\
-1-R
\end{pmatrix},
\quad
{\bf U}(\a +0)  = p_0 T\begin{pmatrix}
Z_\theta^{-1}
\\
-1
\end{pmatrix},
\eeq
The continuity conditions at $x_1=0$ and $x_1=\a$ require the normal velocity $v_1$ and the pressure $p$ to be continuous, that is $
{\bf U}(0+)  = {\bf U}(0-)$ and $
{\bf U}(\a -0)  = {\bf U}(\a +0)$.  
Hence, with $M_{ij} = M_{ij}(\a)$, 
\beq{97}
T\begin{pmatrix}
Z_\theta^{-1}
\\
-1
\end{pmatrix}
=
\begin{pmatrix}
M_{11} & M_{12}  
\\
M_{21} & M_{22} 
\end{pmatrix}
\begin{pmatrix}
Z_\theta^{-1}(1-R)
\\
-1-R
\end{pmatrix} 
\eeq
The transmission and reflection coefficients follow from \eqref{97} as
\bse{56}
\bal{56a}
T &= 2 \big( M_{11} +M_{22} + Z_\theta M_{12} + Z_\theta^{-1} M_{21} 
\big)^{-1} ,
\\
R &= 1 -\big(  M_{22} + Z_\theta M_{12}  \big)\ T .
\eal
\ese
Using the explicit solution for ${\bf M}(\a)$ from \eqref{61} and \eqref{62} yields
\eqref{64}.

\section*{Acknowledgments} Suggestions from the reviewers were helpful. 
 Support under ONR MURI Grant No. N000141310631 is gratefully acknowledged.   

-----------------------------------------
\end{document}